\documentclass[journal]{IEEEtran}
\usepackage{diagbox}
\usepackage{makecell}
\usepackage{algorithm}
\usepackage{algorithmic}
\usepackage{graphicx}
\usepackage{array}
\usepackage{amsmath,amssymb,amsfonts}
\usepackage{cite}
\usepackage{tabularx}
\usepackage{booktabs}
\usepackage{textcomp}
\usepackage{stfloats}
\usepackage{url}
\usepackage{threeparttable}
\usepackage{verbatim}
\usepackage{multirow}
\usepackage{subcaption} % 建议替换为IEEEtran支持的\IEEEsubfigure

\usepackage{enumitem}
\usepackage{etoolbox}
\patchcmd{\IEEEbiography}{\addvspace{\bigskipamount}}{}{}{} 
\usepackage[colorlinks,
            linkcolor=blue,
            anchorcolor=blue,
            citecolor=blue,
            urlcolor=blue
            ]{hyperref}
\begin{document}

% --- 论文标题 ---
\title{Model-Free DRL Control for Power Inverters: From Policy Learning to Real-Time Implementation via Knowledge Distillation}

% --- 作者信息块 (仿照您提供的格式) ---
% 注意：这里我使用了占位符 (First Author 等)，请替换为您实际的作者姓名和单位
\author{ \vskip 1em Yang Yang,  Chenggang Cui, \emph{Member, IEEE},Xitong Niu, \emph{Student Member, IEEE}, \\Jiaming Liu, Chuanlin Zhang, \emph{Senior Member, IEEE}
   
}

% 生成标题
\maketitle
\begin{abstract}
In response to the trade-off between control performance and computational burden hindering the deployment of Deep Reinforcement Learning (DRL) in power inverters, this paper presents a novel model-free control framework leveraging policy distillation. To handle the convergence instability and steady-state errors inherent in model-free agents, an error energy-guided hybrid reward mechanism is established to theoretically constrain the exploration space. More specifically, an adaptive importance weighting mechanism is integrated into the distillation architecture to amplify the significance of fluctuation regions, ensuring high-quality transfer of transient control logic by mitigating the observational bias dominated by steady-state data. This approach efficiently compresses the heavy DRL policy into a lightweight neural network, retaining the desired control performance while overcoming the computational bottleneck during deployment. The proposed method is validated through a hardware-based kilowatt-level experimental platform. Experimental comparison results with traditional methods demonstrate that the proposed technique reduces inference time to the microsecond level and achieves superior transient response speed and parameter robustness.
\end{abstract}


% =========================================================================
% 1. Introduction
% =========================================================================
\section{Introduction}
%With the recent widespread adoption of renewable energy, uninterruptible power supplies (UPS), and electric vehicles, power electronic inverters have become the core cornerstone of modern energy systems, and their control systems directly influence power quality and dynamic stability \cite{Wang2019Harmonic}. Traditional linear controllers, such as PI controllers, rely on linearized models at specific operating points. Consequently, their performance deteriorates sharply under severe transient conditions like sudden load steps, exhibiting sluggish transient response and insufficient robustness \cite{Buso2022Digital}. Furthermore, parameter drift of filter components continuously degrades the optimal operating state of fixed-gain controllers. These inherent limitations have prompted the industry to conduct in-depth exploration into advanced power converter control technologies, broadly forming two dominant research directions: model-based control and data-driven control \cite{Hou2013From}.
Driven by the extensive integration of renewable energy, microgrids, and electric vehicles, power electronic inverters serve as the fundamental interface in modern grids, where the control performance dictates the overall power quality and stability \cite{Wang2019Harmonic}. Traditional linear schemes, typically PI, use linearized models fixed to specific operating points. Under transients such as sudden load steps, performance deteriorates as the response becomes sluggish and robustness suffers. \cite{Buso2022Digital}. Filter parameter drift also compromises static-gain controllers, deviating them from the optimal state. Such limitations drive the industry toward advanced strategies, broadly categorized into model-based and data-driven control \cite{Hou2013From}.
%引言中有句话太短了，不要写UPS、不要用Yet、In the same time太口语化

% In the field of model-based control, a particularly effective paradigm in recent research is the composite control architecture \cite{Chen2016Disturbance}.  \cite{Niu2024New} proposes a feedback stabilization strategy based on a Nonlinear Disturbance Observer (NDO), which addresses the stability assurance problem under complex load conditions without requiring load sensors. \cite{Zhao2021Observer} combines this with Sliding Mode Control (SMC), utilizing the strong robustness of SMC while suppressing disturbances through NDO feedforward compensation. A robust Finite-Time Control (FTC) strategy for three-phase inverters is proposed in \cite{Lin2025Robust}, incorporating an Extended Finite-Time Disturbance Observer (EFDO) to achieve rapid finite-time estimation and suppression of parameter uncertainties and load disturbances. Although these advanced methods have made significant progress, model-based control design relies on precise mechanism models, the establishment of which is not only time-consuming and laborious but also prone to introducing errors.
Composite control architectures currently represent a dominant paradigm in model-based research \cite{Chen2016Disturbance}. To secure stability under complex loads without requiring sensors, a feedback stabilization strategy based on a Nonlinear Disturbance Observer (NDO) is utilized in \cite{Niu2024New}. Further integration with Sliding Mode Control (SMC) leverages SMC's inherent robustness; meanwhile, NDO feedforward compensation actively suppresses disturbances \cite{Zhao2021Observer}. In a similar vein, a robust Finite-Time Control (FTC) solution for three-phase inverters is presented in \cite{Lin2025Robust}. This approach incorporates an Extended Finite-Time Disturbance Observer (EFDO), achieving rapid estimation and suppression of both parameter uncertainties and load disturbances. These control strategies are nonetheless contingent upon precise mechanism models. Deriving these models remains time-consuming, laborious, and prone to error.

% At the same time, power electronics technology is developing towards high frequency, high power density, and complex topologies, increasing the difficulty of inverter design. Consequently, the rigid reliance on precise mathematical models may complicate the controller design process and often necessitates conservative performance margins to guarantee robustness against structural complexities. Moreover, the inherent nonlinearity and time-variance of passive components in power converters can lead to severe parameter mismatches. Such mismatches cause traditional controllers to lose their original phase margin, easily triggering system oscillations or even instability. For instance, the performance of Model Predictive Control (MPC) is highly dependent on the accuracy of the prediction model; when parameter deviation is excessive, its control effect is no longer optimal. Additionally, the tuning of weighting factors often relies on the trial-and-error experience of engineers, making it difficult to guarantee global optimality under all operating conditions \cite{Khalilzadeh2021Model-Free}.
Power electronics technology, simultaneously, evolves toward high frequency, high power density, and complex topologies; this progression increases the difficulty of inverter design. Consequently, controller synthesis is complicated by a rigid reliance on precise mathematical models. Structural complexities often mandate conservative performance margins. Nonlinearity and time-variance in passive components introduce severe parameter mismatches. Phase margin in traditional controllers is easily compromised under these conditions, triggering oscillations or instability. Model Predictive Control (MPC) is strictly governed by the accuracy of the prediction model. Excessive parameter deviation degrades its control effectiveness. Weighting factor tuning relies on engineering trial-and-error, making global optimality elusive \cite{Khalilzadeh2021Model-Free}.

% Consequently, model-free methods have garnered significant attention by circumventing the complexities of precise system modeling. The authors in \cite{Li2022Revisit} constructed a systematic analysis framework for power electronics based on Fliess’s classic MFC theory, providing critical theoretical support for the engineering application of this technology. Furthermore, Model-Free Predictive Control (MFPC) has emerged as a popular solution to mitigate the robustness limitations of Model-Based Predictive Control (MBPC) by substituting physical models with data-driven counterparts. For instance, \cite{Wei2025ModelFree} proposes the ULM-PM-MFPC strategy for PMSM drives; its efficacy in balancing control performance and computational burden offers a valuable reference for the practical implementation of model-free control.
Model-free methods effectively bypass the complexities of precise system modeling. This advantage has spurred extensive research efforts into data-driven alternatives. In \cite{Li2022Revisit}, a systematic analysis framework for power electronics is constructed based on Fliess’s classic MFC theory. Such theoretical work provides critical support, bridging the gap between abstract principles and engineering applications. Meanwhile, Model-Free Predictive Control (MFPC) substitutes rigid physical models with adaptive data-driven counterparts. This substitution strategy mitigates robustness limitations often inherent in Model-Based Predictive Control (MBPC). Specifically, a ULM-PM-MFPC strategy designed for PMSM drives is presented in \cite{Wei2025ModelFree}. The method achieves a trade-off between control performance and computational burden, serving as a practical benchmark for implementation.

The potential of DRL in performance enhancement has been confirmed, achieving significant results in fields such as autonomous driving and robotics \cite{Govinda2025Survey}, and has been introduced into power electronics \cite{She2023Fusion, Mohammadi2022Review, Rajamallaiah2025DRL}. A model-free DRL control strategy based on domain adaptation transfer learning is developed in a recent result \cite{Cui2025Domain} to optimize the system performance of DC microgrids under complex conditions. In \cite{Zhou2024DRL}, the parameter tuning of non-smooth controllers is formulated as a continuous action space problem, where the SAC algorithm is employed to facilitate real-time self-configuration. DRL utilizes Deep Neural Networks (DNN) as universal nonlinear function approximators, capable of directly mapping from raw states to optimal actions and implicitly capturing the complex dynamic characteristics of inverters without any linearization assumptions.  \cite{Wan2025Stability} addresses the stability challenge of Reinforcement Learning (RL) in power converter control by proposing a Lyapunov-guided RL control strategy, which clarifies the actual stability domain of the system by quantifying the convergence compact set of voltage control errors.

Although DRL exhibits excellent performance and robustness in simulation environments, its direct deployment on real power electronics hardware still faces challenges regarding real-time capability and computing power \cite{Li2020EdgeAI}. DRL strategies are typically based on multi-layer neural networks; their complex network structures and massive parameter counts lead to high computational demands, which are incompatible with the stringent real-time requirements of power electronic converters under high switching frequencies and resource-constrained hardware \cite{Rajamallaiah2025DRL}. Secondly, standard DRL reward function designs usually focus only on minimizing instantaneous tracking errors, lacking consideration for the system's long-term dynamic behavior and robustness. This design paradigm tends to guide the agent to converge to suboptimal policies, which may perform excellently under specific conditions but poorly in the face of unknown disturbances \cite{Chen2024RLReview}.

To address the aforementioned challenges, this paper innovatively introduces Policy Distillation technology \cite{Romero2015FitNets,Yang2022Focal}. Policy distillation efficiently transfers the knowledge learned by a complex DRL teacher policy to a compact, computationally efficient lightweight student network. This reduces the model's parameter count and inference latency while maintaining control performance.

The main contributions of this paper are reflected in the following three aspects:
\begin{enumerate}
    \item \textbf{Hybrid Reward Function Design}: A hybrid reward mechanism incorporating a discrete Lyapunov candidate function targets the convergence instability of model-free agents. By \textbf{penalizing error energy increments}, the method constrains exploration within asymptotically stable regions. This design effectively counters suboptimal convergence driven by the greedy optimization of instantaneous returns.
    \item \textbf{Model-Free DRL Control Framework}: High-frequency switching nonlinearities, strong state couplings, and parameter aging complicate mechanistic modeling in inverters. A model-free DRL framework is established to address these specific challenges. DNNs extract latent features directly from raw data, implicitly capturing complex unmodeled dynamics. This capability bypasses reliance on precise mechanistic models.
% \textbf{Policy Distillation}: Capturing complex dynamics demands high model capacity, yet hardware mandates strict latency. A teacher-student policy distillation architecture is introduced to resolve this conflict. The heavy teacher policy is compressed into a lightweight student network. \textbf{Microsecond-level inference is achieved, preserving the robust performance of the complex DRL agent}.
\item \textbf{Policy Distillation}: Capturing complex dynamics demands high model capacity, yet hardware mandates strict latency. A teacher-student policy distillation architecture incorporating adaptive importance weighting is introduced to resolve this conflict. By amplifying the significance of fluctuation regions, this design mitigates the observational bias toward steady-state data, enabling the lightweight student network to achieve microsecond-level inference while effectively preserving the superior transient control performance of the teacher agent.
\end{enumerate}

% The remainder of this paper is organized as follows: Section II establishes the mathematical model of the three-phase inverter, clarifies the control objectives, and introduces the basics of DRL. Section III focuses on the design method of the Lyapunov-guided DRL controller, enhancing policy robustness by constructing a hybrid reward function. Section IV details the controller lightweighting framework based on policy distillation, aiming to solve the deployment problem of complex DRL models on hardware. Section V validates the real-time capability and superior performance of the proposed lightweight controller through simulation and power hardware experiments. Section VI summarizes the research findings of the full paper and looks forward to the future.

 \vspace{-0.5em} 
\section{Problem Formulation }
 \vspace{-0.5em} 
This section first describes the configuration of the studied three-phase Voltage Source Inverter (VSI) system and analyzes the challenges posed by different load characteristics. Subsequently, the mathematical model of the system under load disturbances is established to define the control objectives and core problem formulation of this paper\cite{xin2006engineering}.
 \vspace{-1em} 
\subsection{System Description of the Three-Phase VSI}
\begin{figure}[thpb]
 \vspace{-0.2em} 
  \centering
  \includegraphics[width=1.0\linewidth]{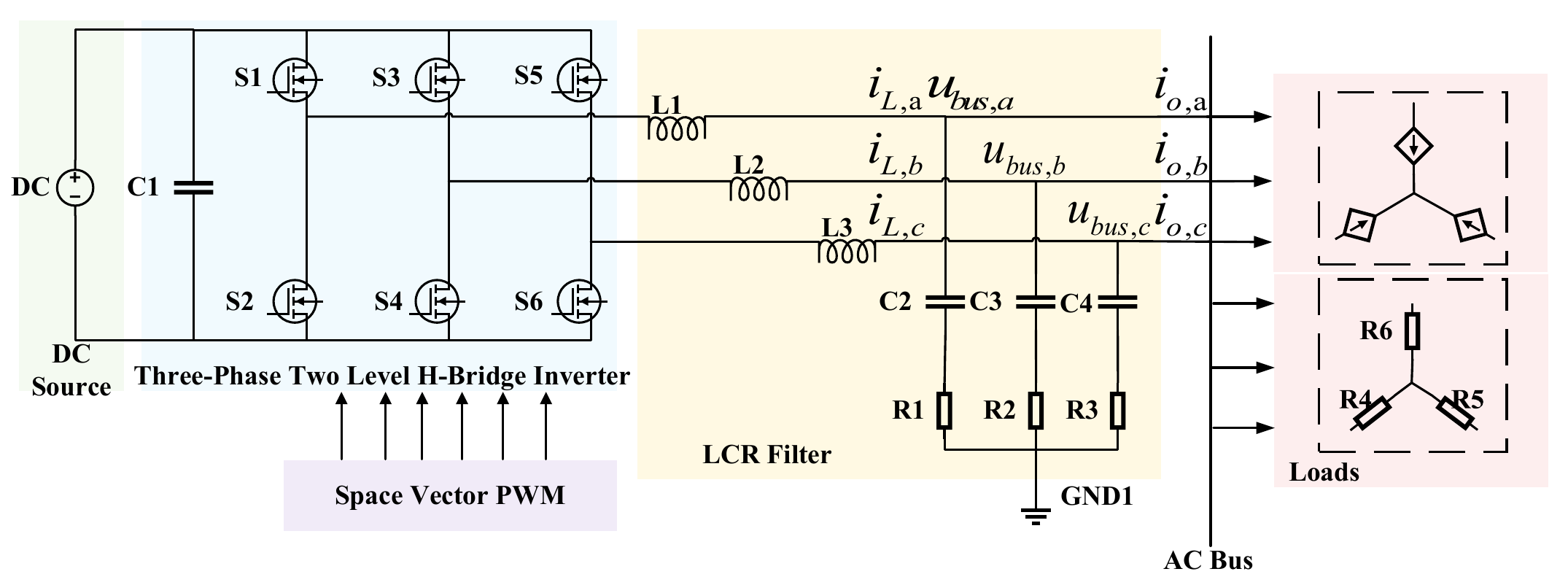} % Placeholder
  \caption{Simplified circuit diagram of an autonomous AC microgrid.}
  \label{fig:framework_architecture_main_v3} 
   \vspace{-0.5em} 
\end{figure}

 \vspace{-0.5em} 
The system under study is a typical three-phase voltage source inverter system, with its topology illustrated in Fig. \ref{fig:framework_architecture_main_v3}. The core of this system is a three-phase two-level VSI, which operates in islanded mode as the main power source to establish the AC bus voltage. The inverter output is connected to a third-order LCR filter to eliminate switching harmonics and supply power to the downstream load.

Kirchhoff's voltage and current laws support the controller design model. Park transformation maps the system from the three-phase stationary frame ($abc$) to synchronous rotating coordinates ($dq$). Dynamic differential equations in the $dq$ frame are derived.
\begin{equation}
\begin{cases}
L_f \frac{di_{Ld}}{dt} = u_{inv,d} - u_{bus,d} + \omega L_f i_{Lq} \\[8pt]
L_f \frac{di_{Lq}}{dt} = u_{inv,q} - u_{bus,q} - \omega L_f i_{Ld} \\[8pt]
mC_f \frac{du_{bus,d}}{dt} = i_{Ld} - i_{od} + \omega mC_f u_{bus,q} \\[8pt]
mC_f \frac{du_{bus,q}}{dt} = i_{Lq} - i_{oq} - \omega mC_f u_{bus,d}
\end{cases}
\end{equation}
where $i_{Ld}, i_{Lq}$ and $u_{bus,d}, u_{bus,q}$ are the $dq$-axis components of the inductor current and bus voltage, respectively; $u_{inv,d}, u_{inv,q}$ are the inverter output voltage control variables; $i_{od}, i_{oq}$ are the load disturbance currents; and $\omega$ is the fundamental angular frequency. The parameter $m$ is a correction coefficient introduced to simplify the third-order filter dynamics. Based on frequency-domain steady-state analysis at the rated angular frequency, it is defined as $m = 1 / \sqrt{1 + \omega^2 C_{f}^2 R^{2}}$. This model reveals the strong coupling and nonlinear characteristics among system state variables, serving as a simulation platform for the subsequent validation of the model-free control strategy.

\noindent\textbf{Remark 1:} The derived mathematical model is employed to construct the simulation environment for offline training. In the implementation phase, the proposed DRL controller utilizes deep architectures to capture complex nonlinearities and coupled dynamics, operating independently of explicit system parameters. Control actions are generated solely based on real-time measurements, characterizing the model-free nature. Ultimately, the model serves merely as the environment for the teacher policy to generate expert demonstrations.

\subsection{Problem Formulation}

As a critical interface in modern power systems, the VSI serves to maintain high-quality AC bus voltage. This paper investigates a stand-alone three-phase VSI system equipped with an LCR filter.

Accurate voltage tracking and robust disturbance rejection constitute the primary control objectives. The controller is formulated to eliminate steady-state error and minimize Total Harmonic Distortion (THD) to comply with power quality standards. Furthermore, the control strategy is required to ensure rapid transient recovery under load steps and preserve stability against parameter perturbations, such as $L_f$ and $C_f$ drift.

\textbf{To quantify this,} the control objective is to regulate the measured AC bus voltage vector $\boldsymbol{u}_{bus} = [u_{bus,d}, u_{bus,q}]^T$ to accurately track the ideal reference trajectory $\boldsymbol{u}_{ref} = [u_{ref,d}, u_{ref,q}]^T$ in the synchronous rotating frame. The instantaneous voltage tracking error vector $\boldsymbol{e}_u(t)$ is defined as:

\begin{equation}
\boldsymbol{e}_u(t) = \boldsymbol{u}_{ref}(t) - \boldsymbol{u}_{bus}(t)
\label{eq:error_vector}
\end{equation}

\noindent where the minimization of the Euclidean norm $||\boldsymbol{e}_u(t)||$ serves as the fundamental basis for the reward formulation in the subsequent DRL framework.

\subsection{Fundamentals of SAC Algorithm}
The voltage control problem is formulated as a Markov Decision Process (MDP). Given the continuous action space of power inverters, the SAC algorithm is adopted~\cite{zeng2025multi, pei2023multitask, oshnoei2024grid}. Unlike standard reinforcement learning, SAC operates on a maximum entropy framework to enhance exploration and robustness against parameter uncertainties. The policy optimization objective is explicitly expressed as:
\begin{multline}
\label{eq:sac_objective}
 J(\theta) = \mathbb{E}_{s_t \sim p_{\text{env}}, a_t \sim \pi_t} \left[ \log \pi_{\theta}(a_t|s_t)Q^{\pi}(s_t, a_t) \right. \\
 \left. - \alpha \text{KL}\left(\pi_{\theta}(\cdot|s_t) \left\| \frac{1}{Z(s_t)} e^{\frac{Q^{\pi}(s_t, \cdot)}{\alpha}}\right.\right) \right]
\end{multline}

\noindent where the objective minimizes the Kullback-Leibler (KL) divergence between the parameterized policy $\pi_{\theta}$ and the energy-based distribution induced by the soft Q-function $Q^{\pi}(s_{t},a_{t})$, and $p_{epv}$ denotes the distribution over initial states. The temperature coefficient $\alpha$ serves as a regularization term, explicitly governing the stochasticity of the agent to balance exploration and exploitation. The term $Z(s_{t})$ acts as the partition function, ensuring the exponential Q-value distribution is properly normalized. This formulation encourages the control policy to align with the optimal Boltzmann distribution while preventing premature convergence. Furthermore, to mitigate Q-value overestimation, a clipped double-Q learning mechanism is incorporated. This design ensures the training stability requisite for high-precision inverter control.

\section{Model-Free DRL Control Strategy for Inverter Systems}

Model uncertainty and disturbances pose significant challenges to traditional control methods. Accordingly, a model-free DRL strategy anchored on the SAC algorithm is proposed. The voltage control task is formulated as an MDP. DNNs are utilized to extract end-to-end mapping relationships directly from interaction data, with the primary objective of maximizing long-term cumulative rewards. This section details the proposed framework configuration, including the definitions of state space, action space, reward mechanism, and network architecture.
 \vspace{-0.2em} 
\subsection{State Space}
The design of the state space is crucial, as it provides the agent with sufficient information to comprehensively evaluate the current operating status. In this study, the state vector $\boldsymbol{s}_t$ is defined as:

\begin{equation}
\boldsymbol{s}_t = [e_{ud}, e_{uq}, u_{bus,d}, u_{bus,q}, i_{Ld}, i_{Lq}]^T
\label{eq:state_space}
\end{equation}

\noindent where $e_{ud} = u_{ref,d} - u_{bus,d}$ and $e_{uq} = u_{ref,q} - u_{bus,q}$ represent the tracking errors in the $dq$ frame; $u_{bus,d}$ and $u_{bus,q}$ denote the actual measured AC bus voltages; and $i_{Ld}$ and $i_{Lq}$ correspond to the measured inductor currents. Before being fed into the neural network, all state variables are normalized to the interval $[-1, 1]$ to enhance training stability and convergence speed.
\subsection{Action Space}
The agent outputs voltage references $a_t = [u_{inv,d}, u_{inv,q}]$ to modulate the inverter and minimize the tracking error $\mathbf{e}(t)$. The action amplitude is restricted by the DC bus voltage $V_{dc}$ and the Space Vector Pulse Width Modulation (SVPWM) linear region limits to avoid hardware overstress. This continuous formulation facilitates decoupled control of the voltage vector components. It also fits the SAC algorithm, enabling efficient policy search in continuous domains. 
\subsection{Deep Neural Network Design}
 
% This study constructs a deep neural network architecture comprising a policy network and a state-action value network. To improve training efficiency and stability, the input states of all networks are processed through linear normalization before entering the network.
Leveraging the universal function approximation of deep neural networks, this study addresses VSI nonlinearities via a DRL framework. The network dimensions are strictly guided by recent RL Scaling Laws \cite{wang2025scaling}, which indicate that SAC performance saturates beyond a depth of four layers but remains highly sensitive to width. Accordingly, a width-dominant architecture is employed. This approach mitigates the computational burden of deep models while ensuring sufficient parameter capacity to prevent underfitting in complex control landscapes.

\textbf{Actor Network:} Responsible for generating control policies. This network consists of an input layer, three fully connected hidden layers, and an output layer. The input layer receives the normalized environmental state. After nonlinear feature extraction through the hidden layers, the output layer branches into two parallel fully connected layers, generating the mean ($\mu$) and standard deviation ($\sigma$) of the action Gaussian distribution, respectively, where the standard deviation path employs a Softplus function to ensure positive definiteness.

\begin{figure}[thpb]
 \vspace{-0.2em} 
  \centering
  \includegraphics[width=1.0\linewidth]{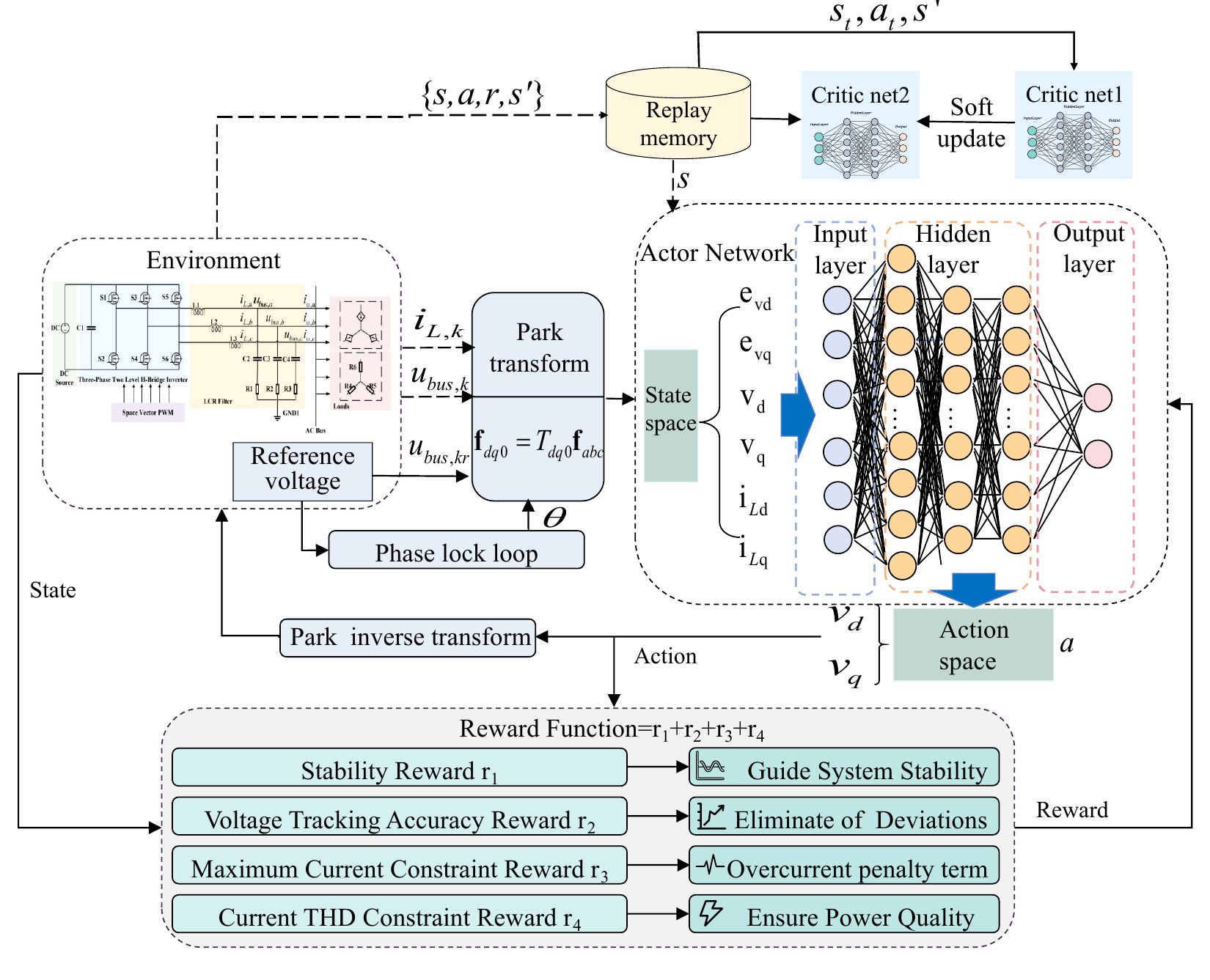} % Placeholder
  \caption{Deep Reinforcement Learning Training Process.}
  \label{fig:framework1} % Updated label
   
\end{figure}

\textbf{Critic Network:} Used to evaluate the value of state-action pairs $Q(s_t, a_t)$. Its structure adopts a dual-input path design: the state path receives the state vector, and the action path receives the action vector. After being processed by fully connected layers respectively, the features are concatenated. The fused features then undergo deep information interaction through subsequent fully connected layers, and finally, a scalar $Q$ value is output by a linear layer. The structure based on SAC algorithm control is shown in Fig. \ref{fig:framework1}.

\subsection{Lyapunov-Based Stability Constraint with Virtual Damping}

The reward function determines the optimization direction of the agent and is crucial for achieving the desired control objectives. In power converter control tasks, traditional reward functions are typically constructed based solely on instantaneous voltage tracking errors. For LCR systems, constraining the voltage error leads to zero-dynamics instability, where the internal inductor current $\mathbf{i}_L$ may still oscillate even when the voltage error approaches zero. This phenomenon poses a severe challenge to the robustness of model-free control strategies.

 To facilitate asymptotic stability, this paper introduces the current increment as a virtual damping term to construct a damping-augmented discrete Lyapunov candidate function $V(k)$, thereby theoretically constraining the exploration space. This design aims to effectively suppress the internal resonance of the LCR filter by penalizing drastic current variations, without requiring prior knowledge of the ideal current reference.

The Lyapunov candidate function $V(k)$, incorporating both voltage tracking error potential energy and internal state damping, is defined as follows:
\begin{equation}
    V(k) = \frac{1}{2} \| \mathbf{e}_u(k) \|^2 + \frac{\beta}{2} \| \Delta \mathbf{i}_L(k) \|^2 
    \label{eq:lyapunov_func}
\end{equation}
where $\mathbf{e}_u(k) = [e_{ud}(k), e_{uq}(k)]^T$ denotes the voltage tracking error vector in the $dq$ reference frame, and $\Delta \mathbf{i}_L(k) = \mathbf{i}_L(k) - \mathbf{i}_L(k-1)$ represents the incremental vector of the inductor current state, with $\mathbf{i}_L(k) = [i_{Ld}(k), i_{Lq}(k)]^T$. The parameter $\beta$ is a positive damping weighting coefficient used to balance tracking accuracy with control smoothness (i.e., vibration suppression).

In the discrete-time domain, the rate of change of this function is approximated as:
\begin{equation}
    \Delta V(k) = V(k) - V(k-1)
    \label{eq:delta_V}
\end{equation}

A positive $\Delta V(k)$ implies an increase in the total system energy, indicating potential system instability. Consequently, a stability-oriented penalty term $r_1$ is constructed to suppress the growth of the Lyapunov function:
\begin{equation}
    r_1(k) = -k_1 \cdot \max(0, \Delta V(k))
    \label{eq:reward_stability}
\end{equation}
where $k_1$ represents a positive penalty weighting factor. The operator $\max(0, \cdot)$ imposes a penalty only when $\Delta V(k) > 0$; when $\Delta V(k) \le 0$, the penalty vanishes, allowing other control objectives to dominate.

Through this mechanism, the agent is guided to learn actions that promote non-increasing system energy. This effectively steers the teacher policy $\pi_T$ toward the stable manifold during the exploration phase, avoiding current spikes and system instability caused by high-frequency switching actions.

% -----------------------------------------------------------------
% Section D: 辅助奖励 - 性能与安全
% -----------------------------------------------------------------
\subsection{Auxiliary Performance and Safety Reward Formulation}

System stability, dynamic performance, and operational safety are balanced. Three auxiliary reward components are integrated with the aforementioned stability constraint after multiple validations.

\subsubsection{Voltage Tracking Accuracy Reward $r_2$}
We design $r_2$ as a quadratic penalty on regulation errors to ensure high-precision tracking. Larger errors trigger stricter penalties. The agent prioritizes resolving voltage deviations, accelerating convergence to the reference. This term corresponds to the negative sum of squared errors in the $dq$ frame, minimizing the deviation between actual and reference voltages. The specific definition is as follows:
\begin{equation}
    r_2 = -k_2(e_{ud}^2 + e_{uq}^2)
\end{equation}

\subsubsection{Maximum Current Constraint Reward $r_3$}
To enforce physical safety constraints within the unconstrained optimization framework of DRL, the reward component $r_3$ is formulated as a soft penalty barrier. This term internalizes the inductor current limit $I_{\text{max}}$ by imposing a quadratic cost on boundary violations. Unlike sparse step-penalties, this continuous formulation provides a dense gradient signal proportional to the violation severity, effectively guiding the policy optimization back to the feasible region. The mathematical expression is defined as:

\begin{equation}
    r_3 = -k_3 \max(0, i_{Ld}^2 + i_{Lq}^2 - I_{\text{max}}^2)
\end{equation}

\noindent where $I_{\text{max}}$ denotes the critical amplitude threshold, and $k_3$ functions as a penalty scaling factor to calibrate the constraint strictness. The $\max(0, \cdot)$ operator activates the penalty strictly during overcurrent excursions, leaving the reward landscape unaffected within the Safe Operating Area (SOA). The quadratic nature of the penalty ensures that large deviations incur substantial costs, thereby accelerating the agent's convergence toward safe control policies.
\subsubsection{Current THD Constraint Reward $r_4$}
The reward component $r_4$ imposes a limit on the total harmonic distortion (THD) to maintain power quality. Exceeding the 5\% threshold triggers an immediate penalty, driving the control strategy to reduce distortion and ensure sinusoidal current waveforms.
\begin{equation}
    r_4 = -k_4 \max(0, (i_{\text{thd}} - 5))
    \label{eq:reward_thd}
\end{equation}
\noindent where $i_{\text{thd}}$ denotes the measured THD value of the output current. $k_4$ functions as a positive weighting coefficient.

In the experimental implementation, the reward components are derived directly from non-normalized physical quantities. Weighting factors undergo scaling to appropriate magnitudes, restricting the total reward to a numerical range essential for SAC value network convergence. Specifically, the coefficients are configured as $k_1=10^{-6}$, $k_2=10^{-3}$, $k_3=10^{-5}$, and $k_4=10^{-4}$. The specific form of the total reward function $r$ is as follows:
\begin{equation}
    \left\{
    \begin{aligned}
        r_1 &= -k_1 \max(0, \Delta V(k)) \\
        r_2 &= -k_2(e_{ud}^2 + e_{uq}^2) \\
        r_3 &= -k_3 \max(0, i_{Ld}^2 + i_{Lq}^2 - I_{\text{max}}^2) \\
        r_4 &= -k_4 \max(0, (i_{\text{thd}} - 5)) \\
        r   &= r_1 + r_2 + r_3 + r_4
    \end{aligned}
    \right.
\end{equation}

\noindent\textbf{Remark 2:} It is pivotal to note that the Deep architecture with sufficient capacity employed is necessitated by the universal function approximation capability required for high-performance VSI control. Given the inverter's strongly coupled dynamics and time-varying parameter uncertainties, a shallow network risks underfitting, failing to capture the intricate nonlinear relationships between system states and optimal actions. Therefore, the teacher agent is designed as a highly capable expert possessing enough representational power to accurately model the complex control boundaries across the full operating range. This design ensures that the teacher policy converges to a global optimum, providing a high-fidelity performance benchmark for the subsequent lightweight student network to mimic during the distillation process.

\section{Algorithm Lightweighting Framework Based on Policy Distillation}

Although DRL agents exhibit decent performance, their complex network structures conflict with the  requirements for computational resources and real-time capability in power electronics control systems. To resolve the contradiction between high performance and limited computing power, this section proposes a lightweight framework based on policy distillation.
\begin{figure}[thpb]
 \vspace{-0.2em} 
  \centering
  \includegraphics[width=0.95\linewidth]{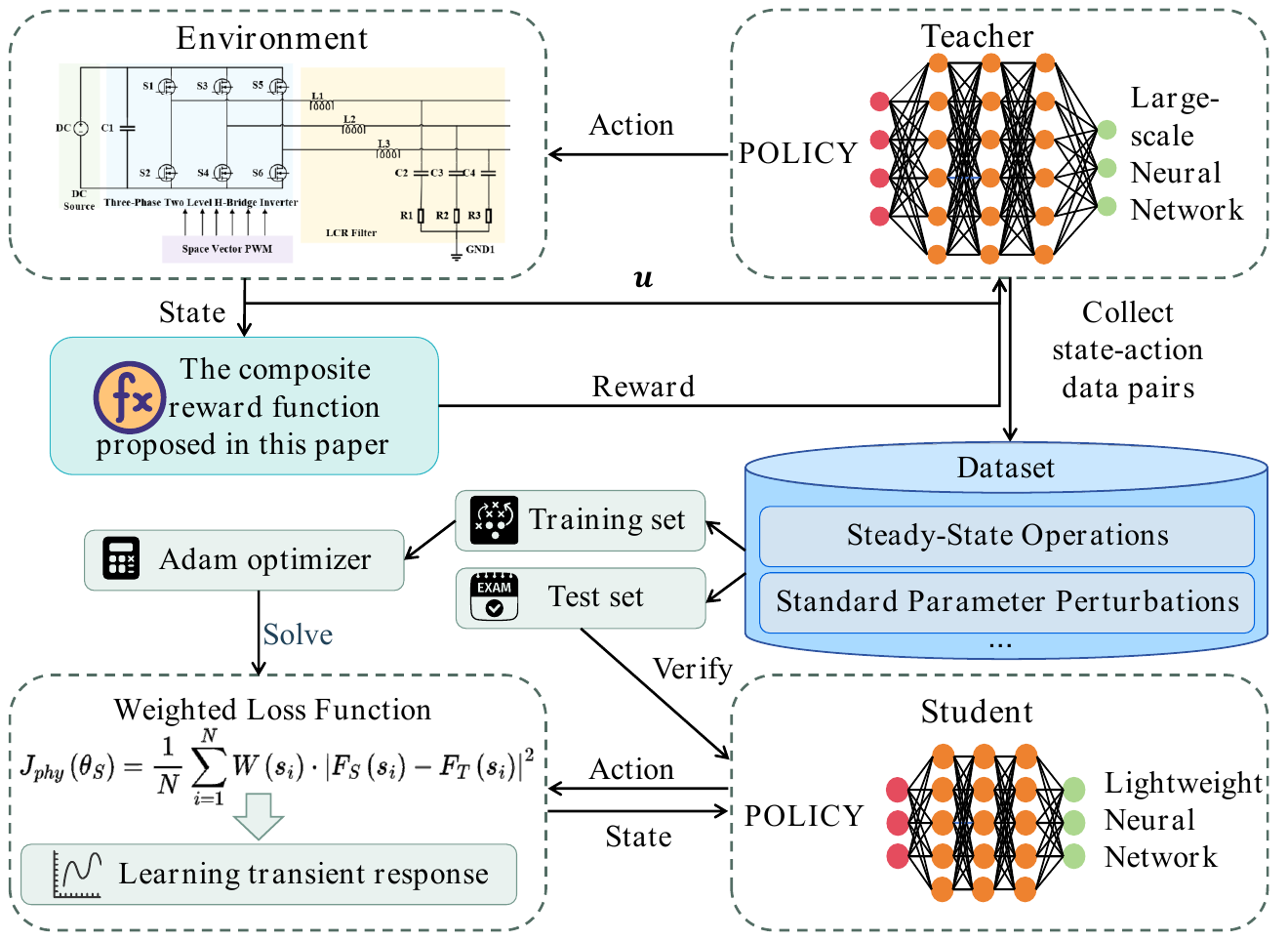} % Placeholder
  \caption{Design process of Policy Distillation.}
  \label{fig:framework_diagram} 
    \vspace{-0.5em} 
\end{figure}
\subsection{Principle of Policy Distillation}

The idea of policy distillation is to construct a teacher-student learning paradigm to achieve knowledge transfer from a complex model to a lightweight model, while maintaining its original performance as much as possible. In the context of DRL, this is not merely a simple parameter reduction, but a transfer process involving function approximation and generalization capabilities.

The trained teacher policy is defined as the mapping function of the source domain, and the student policy to be deployed is defined as the mapping function of the Target Domain. The distillation process essentially seeks the optimal solution for the student network parameters within the state space distribution, making its decision boundary approach that of the teacher network infinitely.

\subsection{Teacher-Student Knowledge Transfer Paradigm}

Referring to the application of transfer learning in control systems \cite{Cui2025DomainAdaptation}, we formulate the policy distillation process as a function approximation problem constrained by model complexity.

First, the Teacher Policy is defined as a deep neural network $\pi_{T}$ parameterized by $\theta_{T}$. To explicitly represent its deep architecture capable of extracting hierarchical features, the mapping from the high-dimensional state space $s \in \mathcal{S}$ to the action space is expressed as a deep composite function $\mathcal{F}_{T}$:
\begin{equation}
\label{eq:teacher_policy}
\begin{split}
    a_{T} &= \pi_{T}(s;\theta_{T}) = \mathcal{F}_{T}(s; \theta_{T}) \\
          &= \phi^{(L)}(\dots \phi^{(1)}(s; \mathbf{W}_{T}^{(1)}, \mathbf{b}_{T}^{(1)})\dots; \mathbf{W}_{T}^{(L)}, \mathbf{b}_{T}^{(L)})
\end{split}
\end{equation}
where $\phi^{(l)}(\cdot)$ denotes the non-linear activation function of the $l$-th layer, and $\theta_{T} = \{\mathbf{W}_{T}^{(l)}, \mathbf{b}_{T}^{(l)}\}_{l=1}^{L}$ represents the massive set of weights and biases in the teacher network. This deep composite structure $\mathcal{F}_{T}$ enables the teacher to capture the complex, highly coupled dynamics of the inverter.

Correspondingly, the Student Policy $\pi_{S}$, parameterized by $\theta_{S}$, is designed as a shallow or compact network to satisfy hardware constraints. Its output action $a_{S}$ is defined as:
\begin{equation}
\label{eq:student_policy}
a_{S} = \pi_{S}(s;\theta_{S}) = \mathcal{F}_{S}(s; \theta_{S})
\end{equation}
% \noindent where the action variable $a$, representing either the teacher output $a_T$ or the student output $a_S$, corresponds to the inverter control voltage vector $u_{inv}$ described in Section II. The key constraint for hardware deployment is $|\theta_{S}|\ll|\theta_{T}|$. The goal of policy distillation is to find the optimal student parameters $\theta_{S}^{*}$ that minimize the divergence between the student's simplified mapping $\mathcal{F}_{S}$ and the teacher's expert mapping $\mathcal{F}_{T}$ over the state distribution $\rho(s)$ induced by the teacher.

% The optimization objective is formulated as minimizing the expected loss:
% \begin{equation}
% \label{eq:optimization_obj}
% \theta_{S}^{*} = \arg \min_{\theta_{S}} \mathbb{E}_{s \sim \rho(s)} [\mathcal{L}(\mathcal{F}_{T}(s;\theta_{T}), \mathcal{F}_{S}(s;\theta_{S}))]
% \end{equation}

\noindent where the action variable $a$, representing either the teacher output $a_T$ or the student output $a_S$, corresponds to the inverter control voltage vector $u_{inv}$ described in Section II. The key constraint for hardware deployment is $|\theta_{S}|\ll|\theta_{T}|$. To achieve efficient knowledge transfer under this constraint, the optimization objective of policy distillation is formulated as:

\begin{equation}
    \theta_{S}^{*} = \arg \min_{\theta_{S}}\mathbb{E}_{s\sim\rho(s)}[\mathcal{L}(\mathcal{F}_{T}(s;\theta_{T}),\mathcal{F}_{S}(s;\theta_{S}))]
    \label{eq:general_objective}
\end{equation}

\noindent where $\theta_{S}^{*}$ denotes the optimal student parameters. This formulation minimizes the expected divergence $\mathcal{L}$ between the expert mapping $\mathcal{F}_{T}$ and the simplified mapping $\mathcal{F}_{S}$ over the distribution $\rho(s)$. Consequently, the student network accurately reconstructs the teacher's decision boundary within a compact parameter space.-

\subsection{Distillation with Adaptive Importance Weighting and Lyapunov Consistency}

While standard MSE minimization is valid for continuous action regression, its application in power electronics faces two critical limitations. First, the overwhelming dominance of steady-state data induces severe observational bias \cite{lou2021physics}, causing the student network to prioritize abundant steady-state behaviors at the expense of sparse yet crucial transient dynamics. Second, to deploy the lightweight student network $\pi_S$ while retaining the stability characteristics inherited from the teacher policy, the distillation process requires explicit regularization beyond simple action matching. To address these dual challenges, we propose a Lyapunov-Preserved Policy Distillation framework. This method constructs a composite loss function $J_{phy}(\theta_{S})$ that integrates an adaptive importance weight $\mathcal{W}(s_{k})$ with a Lyapunov-consistency constraint $\mathcal{L}_{stab}$, mathematically defined as:

\begin{equation}
\begin{split}
    J_{phy}(\theta_{S}) = \frac{1}{N}\sum_{k=1}^{N}\mathcal{W}(s_{k}) \cdot \Big( & \left\|a_{T,k}-\mathcal{F}_{S}(s_{k};\theta_{S})\right\|_{2}^{2} \\
    & + \lambda_{stab}\cdot \max(0,\Delta V(s_{k}, a_{S,k})) \Big)
\end{split}
\label{eq:distillation_loss}
\end{equation}

where $N$ denotes the mini-batch size, $s_k$ represents the $k$-th state sample in the batch, $a_{T,k}$ is the corresponding target action from the teacher, and $\lambda_{stab}$ is the weighting coefficient. To specifically mitigate observational bias, the gradient-guided dynamic penalty factor $\mathcal{W}(s_{k})$ is formulated as:

\begin{equation}
    \mathcal{W}(s_{k})=1+\omega_{p}\cdot\mathbb{I}(\left\|e_{u}(k)-e_{u}(k-1)\right\|_{2}>\delta_{th})
    \label{eq:adaptive_weight}
\end{equation}

where $\mathbb{I}(\cdot)$ is the indicator function. This weight imposes a higher penalty gain $\omega_{p}$ on samples where the transient error change rate exceeds a threshold $\delta_{th}$. Simultaneously, the Lyapunov-consistency term introduces a stability regularization derived from the discrete Lyapunov candidate function $V(\cdot)$, where $\Delta V(s_{k}, a_{S,k})$ represents the predicted energy change. The operator $\max(0, \cdot)$ activates a penalty only when the student's action leads to an increase in the Lyapunov function. By explicitly penalizing control actions that result in a positive increment of the Lyapunov function, this mechanism facilitates the lightweight student policy in complying with the asymptotic stability requirements.

\subsection{Dataset Construction and Supervised Training}

To ensure that that the student network generalizes to unknown disturbances rather than memorizing specific state points, this paper adopts a  \textit{trajectory-based data partition strategy}. Instead of random sampling, the expert dataset $\mathcal{D}_{expert}$ is constructed by collecting complete time-series trajectories generated by the teacher agent under closed-loop control. The dataset consists of $M$ independent trajectories, mathematically expressed as:

\begin{equation}
    \mathcal{D}_{expert} = \{ \tau_i \}_{i=1}^{M}, \quad \text{where } \tau_i = \{ (s_t, a_t) \}_{t=0}^{T_i}
    \label{eq:dataset_trajectory}
\end{equation}

\noindent where $\tau_i$ represents the $i$-th trajectory with a duration of $T_i$, containing sequential state-action pairs. Crucially, we enforce a strict isolation between training and testing data at the trajectory level. The training set $\mathcal{D}_{train}$ covers steady-state operations and standard parameter perturbations, while the test set $\mathcal{D}_{test}$ is reserved for a completely unseen load step experiment that never appears during training. This design rigorously validates the student policy's capability to handle unknown transient dynamics.

Under this framework, the training of the student network is transformed into a supervised learning process on $\mathcal{D}_{train}$. By using the Adam optimizer to minimize the weighted distillation loss defined in \eqref{eq:distillation_loss}, the student network inherits the teacher's robust control capability while significantly compressing the computational load. The finally obtained lightweight mapping function $\mathcal{F}_S(\cdot)$ is then compiled and deployed to hardware for millisecond-level real-time control. Fig. \ref{fig:framework_diagram} illustrates the entire process from DRL control to Algorithm Lightweighting.

% \begin{figure*}[thpb]
%  \vspace{-0.2em} 
%   \centering
%   \includegraphics[width=0.85\linewidth]{paper3/framework.pdf} % Placeholder
%   \caption{Design process of the proposed controller.}
%   \label{fig:framework_diagram} 
%     \vspace{-0.5em} 
% \end{figure*}

\noindent\textbf{Remark 3:} It is worth emphasizing that the proposed policy distillation framework fundamentally transforms the computationally intensive stochastic policy inference into a streamlined supervised regression task. Leveraging the highly capable expert policy established in Section III, the lightweight student network is trained to accurately reconstruct the optimized policy mapping within a significantly compact parameter space. This paradigm allows the controller to inherit the robustness derived from the teacher's deep feature extraction and generalization capabilities, while bypassing the heavy computational burden, thus effectively bridging the gap between advanced AI capabilities and the stringent latency constraints of power electronic hardware.
\section{Simulation and Experimental Validation}
\label{sec:validation}
 \vspace{-0.5em} 
Simulations and rapid control prototyping experiments validate the proposed DRL framework. The controller is benchmarked against conventional strategies to reveal performance differences. Key evaluation metrics include dynamic response, steady-state accuracy, and robustness. Feasibility in physical systems is confirmed through these tests.
 \vspace{-0.2em} 
\subsection{Simulation Verification}
\label{sec:simulation}
 \vspace{-0.5em} 
% A simulation model of the three-phase voltage source inverter is constructed as shown in Fig. \ref{fig:framework_architecture_main_v3}. Table \ref{tab:params} lists the circuit and controller parameters, including the specific hyperparameters configured for the policy distillation process described in Section \ref{sec:iv}. Various disturbances are applied to test transient response and steady-state recovery. These scenarios provide quantitative metrics for performance comparison.
A simulation model of the three-phase voltage source inverter is constructed as shown in Fig. \ref{fig:framework_architecture_main_v3}. Table \ref{tab:circuit_params} lists the main circuit specifications. The control parameters, including the specific hyperparameters configured for the policy distillation process, are detailed in Table \ref{tab:drl_params}. Various disturbances are applied to test transient response and steady-state recovery. These scenarios provide quantitative metrics for performance comparison.
\begin{table}[htbp]
\caption{Main Circuit Parameters of VSI}
\label{tab:circuit_params}
\centering
% 增加行高，使表格看起来更大气、清晰
\renewcommand{\arraystretch}{1.3}
\begin{tabular}{lcl}
\hline
\textbf{Parameter} & \textbf{Symbol} & \textbf{Value} \\
\hline
DC source voltage & $V_{dc}$ & 650 V \\
Nominal line voltage & $V_{line}$ & 380 V \\
DC-link capacitance & $C_{1}$ & 2000 $\mu$F \\
Filter inductance & $L_{f}$ & 1.1 mH \\
Filter capacitance & $C_{f}$ & 19.2 $\mu$F \\
Damping resistance & $R_{f}$ & 3 $\Omega$ \\
Switching frequency & $f_{sw}$ & 10 kHz \\
\hline
\end{tabular}
\end{table}

\begin{table}[htbp]
\caption{Hyperparameters for DRL and Policy Distillation}
\label{tab:drl_params}
\centering
% 同样增加行高
\renewcommand{\arraystretch}{1.3}
\begin{tabular}{lcl}
\hline
\textbf{Parameter} & \textbf{Symbol} & \textbf{Value} \\
\hline
Episode number & $M$ & 500 \\
Discount factor & $\gamma$ & 0.9 \\
Learning rate & $\eta$ & $1\times 10^{-3}$ \\
Mini-batch size & $B$ & 256 \\
Sampling time & $T_{s}$ & $1\times 10^{-4}$ s \\
Target Entropy & $T_{e}$ & -13 \\
Transient detection threshold & $\delta_{th}$ &  0.0048 V \\
Distillation penalty gain & $\omega_{p}$ & 9.0 \\
\hline
\end{tabular}
\end{table}
% \begin{table}[htbp]
% \caption{DRL and Main Circuit Parameters}
% \begin{center}
% \begin{tabular}{lcl}
% \hline
% \textbf{Parameter} & \textbf{Symbol} & \textbf{Value} \\
% \hline
% DC source voltage & $V_{dc}$ & 650V \\
% Nominal line voltage & $V_{line}$ & 380V \\
% DC-link capacitance & $C_{1}$ & 2000$\mu$F \\
% Filter inductance & $L_{f}$ & 1.1mH \\
% Filter capacitance & $C_{f}$ & 19.2$\mu$F \\
% Damping resistance & $R_{f}$ & 3$\Omega$ \\
% Switching frequency & $f_{sw}$ & 10kHz \\
% Episode number & $M$ & 500 \\
% Discount factor & $\gamma$ & 0.9 \\
% Learning rate & $\eta$ & 1e-3 \\
% Mini-batch size & $B$ & 256 \\
% Sampling time & $T_{s}$ & 1e-4 \\
% Target Entropy & $T_{e}$ & -13 \\
% Transient detection threshold & $\delta_{th}$ & 1.5V \\  % 新增
% Distillation penalty gain & $\omega_{p}$ & 9.0 \\    % 新增
% \hline
% \end{tabular}
% \label{tab:params}
% \end{center}
% \end{table}
%这个表分开，大一些，然后一些性能指标在实验中去说明。
Four established strategies benchmark the proposed DRL framework. The primary network utilizes three hidden layers (128, 64, 64 neurons). External comparisons include standard dual-loop PI~\cite{monshizadeh2022nonlinear} and Finite Control Set MPC~\cite{cortes2009model}. Internal comparisons involve two distilled variants: S1 (45, 30, 30) and the compact S2 (20, 15). The control efficacy is quantified via SSE, THD, and Relative Overshoot.
\begin{figure}[htbp]
    \centering 
    % 子图1：宽度设为0.8\linewidth（不满行宽，减少水平/垂直占用）
    \begin{subfigure}[b]{0.9\linewidth}
        \centering
        \includegraphics[width=\linewidth]{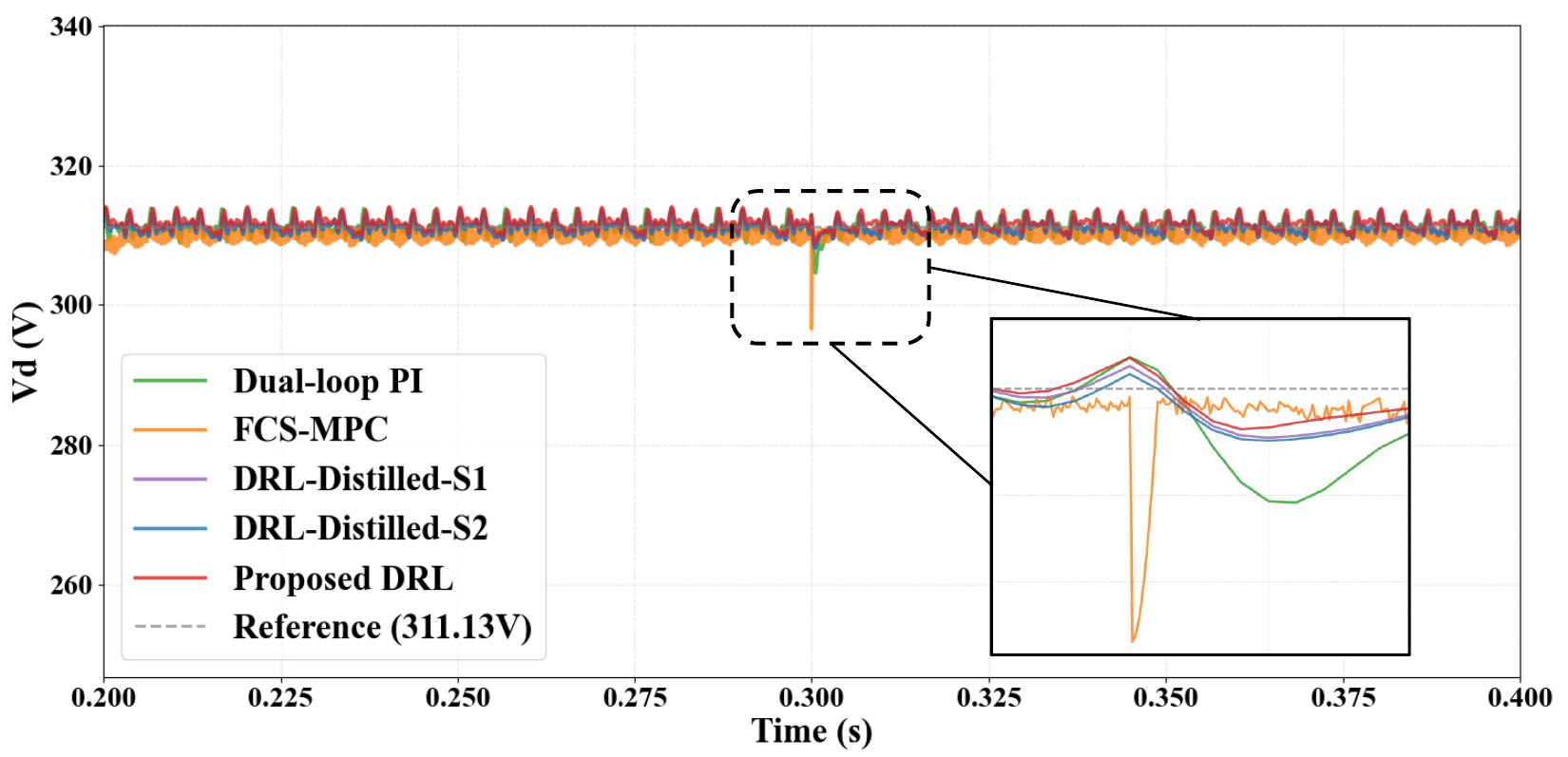} 
        \caption{Case 1: Severe Linear Load Step (R steps from 200$\Omega$ to 50$\Omega$).}
        \label{fig:case1_a}
    \end{subfigure}
    \vspace{0.2em} % 压缩子图间距（原0.5em，减少冗余）
    
    % 子图2：同宽设置
    \begin{subfigure}[b]{0.9\linewidth}
        \centering
        \includegraphics[width=\linewidth]{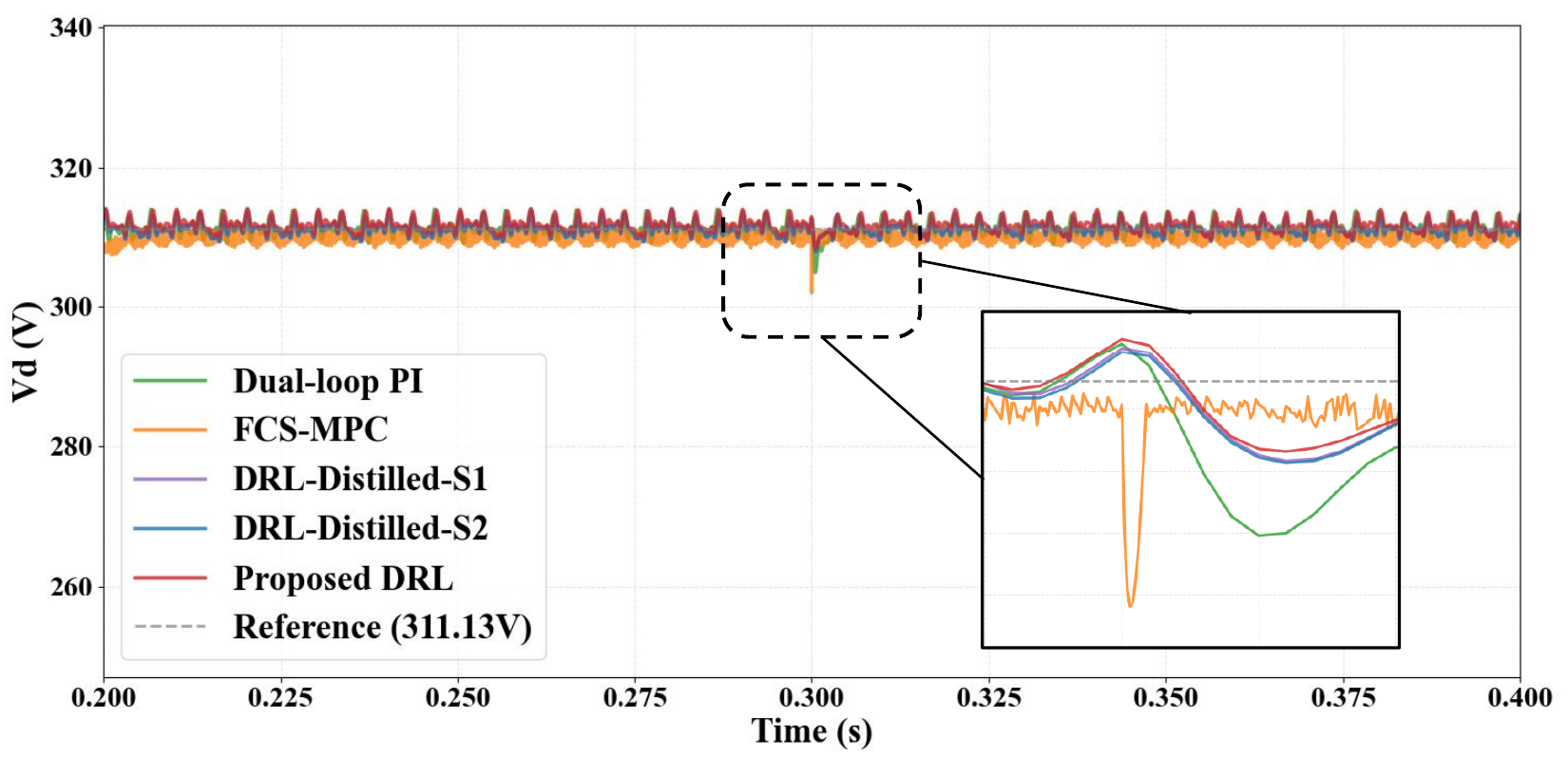} 
        \caption{Case 2: Complex RL Load Switching (RL steps from 200$\Omega$ to 50$\Omega$).}
        \label{fig:case1_b}
    \end{subfigure}
    \vspace{0.2em} % 压缩子图间距
    
    % 子图3：同宽设置
    \begin{subfigure}[b]{0.9\linewidth}
        \centering
        \includegraphics[width=\linewidth]{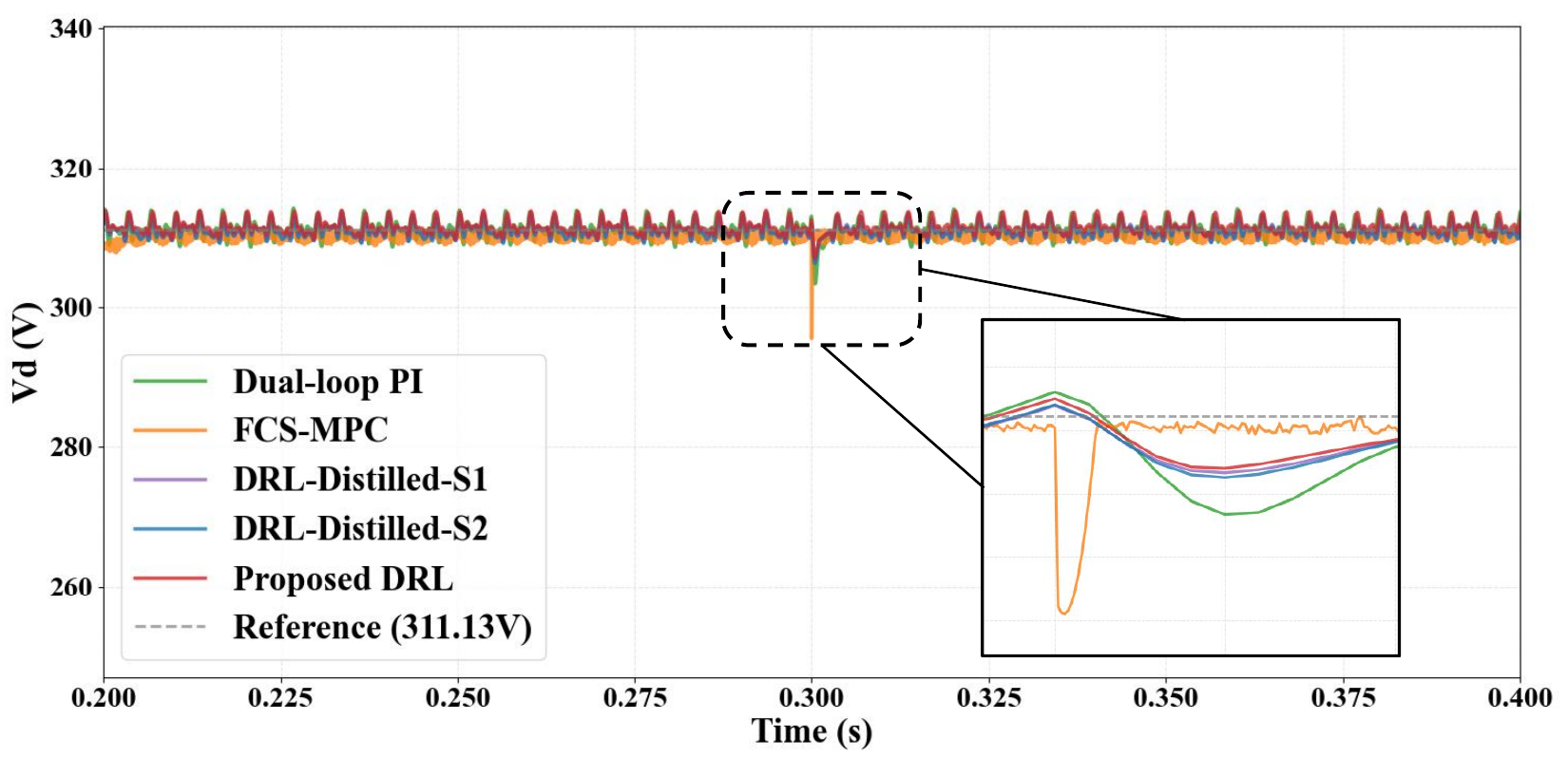} 
        \caption{Case 3: Linear Load Step under Parameter Uncertainty ($R$ steps from $200\,\Omega$ to $50\,\Omega$ with $L_f+20\%$ and $C_f-20\%$).}
        \label{fig:case1_c}
    \end{subfigure}

    \caption{Dynamic response comparison of the five controllers under different simulation scenarios}
    \label{fig:all_cases} 
    \vspace{-0.6em} % 进一步压缩图下方空白
\end{figure}

% \begin{figure}[thpb]
%     \centering 
%     \begin{subfigure}[b]{1\textwidth}
%         \centering
%         \includegraphics[width=\linewidth]{images/case1.pdf} 
%         \caption{Case 1: Severe Linear Load Step (R steps from 200$\Omega$ to 50$\Omega$).}
%         \label{fig:case1_a}
%     \end{subfigure}
    
%     \begin{subfigure}[b]{1\textwidth}
%         \centering
%         \includegraphics[width=\linewidth]{images/case2.pdf} 
%         \caption{Case 2: Complex RL Load Switching (RL steps from 200$\Omega$ to 50$\Omega$).}
%         \label{fig:case1_b}
%     \end{subfigure}
    
%     \begin{subfigure}[b]{1\textwidth}
%         \centering
%         \includegraphics[width=\linewidth]{images/case3.pdf} 
%         \caption{Case 3: Load Step under Parameter Uncertainty (R steps from 200$\Omega$ to 50$\Omega$ with L+20\%, C-20\%).}
%         \label{fig:case1_c}
%     \end{subfigure}

%     \caption{Dynamic response comparison of the five controllers under different simulation scenarios (Three-phase output voltage $v_{abc}$ and d-axis voltage $v_d$).}
%     \label{fig:all_cases} 
% \end{figure}
\begin{table}[htbp]
  \centering
  \small
  \renewcommand{\arraystretch}{0.95} 
  \setlength{\tabcolsep}{3.5pt}      
  \caption{PERFORMANCE METRICS OF DIFFERENT METHODS UNDER CASE 1}
  \label{tab:case1}
  \begin{tabular}{@{}lccc@{}}
    \toprule
    \makecell[c]{\textbf{Method}} & 
    \makecell[c]{\textbf{SSE}\\ \textbf{(V)}} & 
    \makecell[c]{\textbf{THD}\\ \textbf{(\%)}} & 
    \makecell[c]{\textbf{Relative Overshoot}\\ \textbf{(\%)}} \\
    \midrule
     Dual-loop PI & 0.01 & 1.45 & 2.11 \\
     FCS-MPC & 1.2 & 0.18 & 4.69 \\
     DRL-Distilled-S1 & 0.06 & 1.23 & 0.91 \\
     DRL-Distilled-S2 & 0.49 & 1.24 & 0.97 \\
     Proposed DRL & 0.05 & 1.15 & 0.84 \\
    \bottomrule
  \end{tabular}
\end{table}
\subsubsection{Case 1: Severe Linear Load Step}
The condition is set such that at $t = 0.3$\,s, the purely resistive load undergoes a step decrease from $200\,\Omega$ to $50\,\Omega$.

As illustrated in Fig.~\ref{fig:all_cases}(a) and Table~\ref{tab:case1}, all DRL-based controllers exhibit minimal overshoot. Specifically, the overshoot of the proposed controller is limited to 0.84\%, which is superior to the 2.11\% of the PI controller and the 4.69\% of the FCS-MPC. Although FCS-MPC achieves the lowest THD of 0.18\%, it suffers from a large steady-state error (SSE) of 1.2\,V and significant transient fluctuations. Consequently, under purely resistive load step conditions, the proposed DRL controller achieves the optimal balance between dynamic response speed and steady-state accuracy in comparison with PI and FCS-MPC.

\begin{table}[htbp] 
  \centering
  \small
  \renewcommand{\arraystretch}{0.95}
  \setlength{\tabcolsep}{3.5pt}
  \caption{PERFORMANCE METRICS OF DIFFERENT METHODS UNDER CASE 2}
  \label{tab:case2} 
  \begin{tabular}{@{}lccc@{}}
    \toprule
    \makecell[c]{\textbf{Method}} & 
    \makecell[c]{\textbf{SSE}\\ \textbf{(V)}} & 
    \makecell[c]{\textbf{THD}\\ \textbf{(\%)}} & 
    \makecell[c]{\textbf{Relative Overshoot}\\ \textbf{(\%)}} \\
    \midrule
     Dual-loop PI & 0.01 & 1.42 & 2.01 \\
     FCS-MPC & 0.41 & 0.19 & 2.95 \\
     DRL-Distilled-S1 & 0.08 & 1.24 & 1.04 \\
     DRL-Distilled-S2 & 0.51 & 1.24 & 1.07 \\
     Proposed DRL & 0.05 & 1.18 & 0.92 \\
    \bottomrule
  \end{tabular}
    \vspace{-0.5em} 
\end{table}
\subsubsection{Case 2: Complex RL Load Switching}
The experimental condition is established such that at $t = 0.3$\,s, the load switches from an initial state of $200\,\Omega$ and $10$\,mH to a final state of $50\,\Omega$ and $1$\,mH.

The corresponding quantitative results are listed in Table~\ref{tab:case2}. The proposed controller and its distilled variants demonstrate superior damping, restricting the overshoot to below 1\%. In contrast, the PI controller exhibits a slower response with a higher overshoot of 2.01\%. While the FCS-MPC method maintains a competitive THD performance, its voltage overshoot reaches 2.95\% and is accompanied by a steady-state error of 0.41\,V. Consequently, under resistive-inductive load switching conditions, the proposed DRL controller outperforms the traditional PI and FCS-MPC schemes in terms of overall dynamic and steady-state performance.

% \begin{table}[htbp] 
%   \centering
%   \small
%   \renewcommand{\arraystretch}{0.95}
%   \setlength{\tabcolsep}{3.5pt}
%   \caption{PERFORMANCE METRICS OF DIFFERENT METHODS UNDER CASE 2}
%   \label{tab:case2} 
%   \begin{tabular}{@{}lccc@{}}
%     \toprule
%     \makecell[c]{\textbf{Method}} & 
%     \makecell[c]{\textbf{$|SSE|$}\\ \textbf{(V)}} & 
%     \makecell[c]{\textbf{THD}\\ \textbf{(\%)}} & 
%     \makecell[c]{\textbf{Relative Overshoot}\\ \textbf{(\%)}} \\
%     \midrule
%      Dual-loop PI & 0.01 & 1.42 & 2.01 \\
%      FCS-MPC & 0.41 & 0.19 & 2.95 \\
%      DRL-Distilled-S1 & 0.05 & 1.24 & 1.07 \\
%      DRL-Distilled-S2 & 0.08 & 1.24 & 1.04 \\
%      Proposed DRL & 0.51 & 1.18 & 0.92 \\
%     \bottomrule
%   \end{tabular}
%     \vspace{-0.5em} 
% \end{table}

\subsubsection{Case 3: Load Step under Parameter Uncertainty}
To evaluate robustness, the system is tested under parameter mismatch conditions where $L_f$ is increased by 20\% and $C_f$ is decreased by 20\%. Under this configuration, a load step change from $100\,\Omega$ to $50\,\Omega$ is applied at $t = 0.3$\,s.

PI and FCS-MPC degrade (Fig.~\ref{fig:all_cases}(c), Table~\ref{tab:case3}). FCS-MPC overshoot spikes to 5.02\%. Parameter deviations are tolerated by DRL variants. The proposed controller limits overshoot to 1.33\%. Robustness against mismatch surpasses conventional benchmarks.

\begin{table}[htbp]
  \centering
  \small                 
  \renewcommand{\arraystretch}{0.95} 
  \setlength{\tabcolsep}{3.5pt}      
  \caption{PERFORMANCE METRICS OF DIFFERENT METHODS UNDER CASE 3}
  \label{tab:case3}
  \begin{tabular}{@{}lccc@{}}
    \toprule
    \makecell[c]{\textbf{Method}} & 
    \makecell[c]{\textbf{SSE}\\ \textbf{(V)}} & 
    \makecell[c]{\textbf{THD}\\ \textbf{(\%)}} & 
    \makecell[c]{\textbf{Relative Overshoot}\\ \textbf{(\%)}} \\
    \midrule
     Dual-loop PI & 0.01 & 1.56 & 2.50 \\
     FCS-MPC & 1.00 & 0.15 & 5.02 \\
     DRL-Distilled-S1 & 0.12 & 1.39 & 1.44 \\
     DRL-Distilled-S2 & 0.41 & 1.42 & 1.56 \\
     Proposed DRL & 0.05 & 1.33 & 1.33 \\
    \bottomrule
  \end{tabular}
\end{table}
 
\subsubsection{Policy Distillation and Quantitative Analysis}
% The ablation comparison of the complexity between the teacher and student networks is shown in Table \ref{tab:distill}. The parameter count of the proposed DRL (teacher) is 13442. The compression ratio for DRL-Distilled-S1 is 5 times; for DRL-Distilled-S2, it reaches 26.7 times. The distillation loss for both remains at a low level. The policy distillation framework significantly reduces the computational complexity of the model while effectively retaining the core performance of the control strategy, verifying the feasibility of Algorithm Lightweighting.
Table \ref{tab:distill} details the network complexity ablation. Inference FLOPs measure single-pass workload. A standard TMS320F28379D DSP with 800 MFLOPS throughput determines theoretical execution time. The teacher policy holds 13,442 parameters. This requires 26.4 k FLOPs or 33.0 $\mu$s. Distillation reduces the load. S1 and S2 models achieve compression ratios of 5 and 26.7. S2 cost drops to 0.9 k FLOPs. Inference takes 1.1 $\mu$s. This duration occupies 1.1\% of the 10 kHz control cycle. Lightweight deployment feasibility is confirmed.

% \begin{table}[htbp]
%   \centering
%   \caption{Quantitative Ablation Comparison of Policy Distillation}
%   \label{tab:distill}
%   % 1. 将 lcccc 改为 lccc (减少了一列)
%   \begin{tabular}{@{}lccc@{}}
%     \toprule
%     % 2. 删除了 Network Structure 的表头
%     \textbf{Method} & 
%     \textbf{Parameters} & 
%     \makecell{\textbf{Compression} \\ \textbf{Ratio}} & 
%     \makecell{\textbf{Distillation} \\ \textbf{Loss}} \\
%     \midrule
%     % 3. 删除了每行对应的结构数据 (如 128/64/64 等)
%     Proposed DRL & 13442 & 1 & / \\
%     DRL-Distilled-S1 & 2687 & 5 & 0.025/0.069 \\
%     DRL-Distilled-S2 & 487 & 26.7 & 0.024/0.079 \\
%     \bottomrule
%   \end{tabular}
% \end{table}
\begin{figure*}[htbp] % 使用 figure* 以跨越双栏
    \centering
    \captionsetup[subfigure]{labelformat=parens, labelsep=quad, skip=1pt, font=small} 
    
    % --- 第一行 (a) 到 (d) ---
    \begin{subfigure}[b]{0.245\linewidth} % 宽度设为约 1/4
        \centering
        \includegraphics[width=\textwidth]{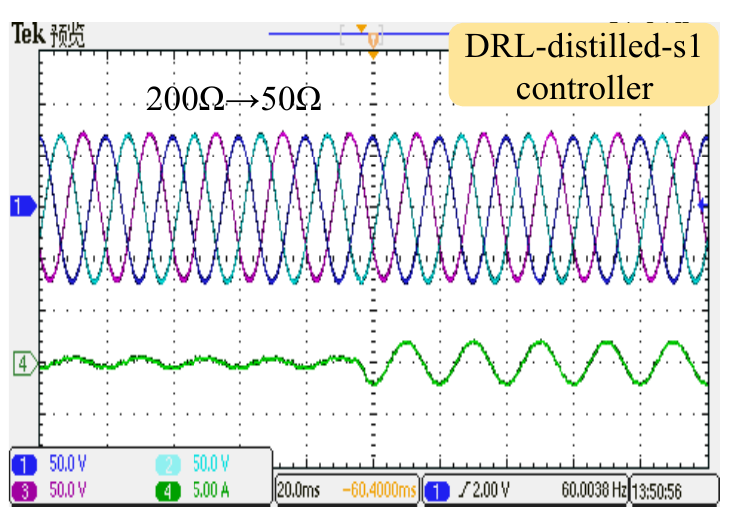} 
       \caption{}
        \label{fig:exp:a} 
    \end{subfigure}
    \hfill 
    \begin{subfigure}[b]{0.245\linewidth}
        \centering
        \includegraphics[width=\textwidth]{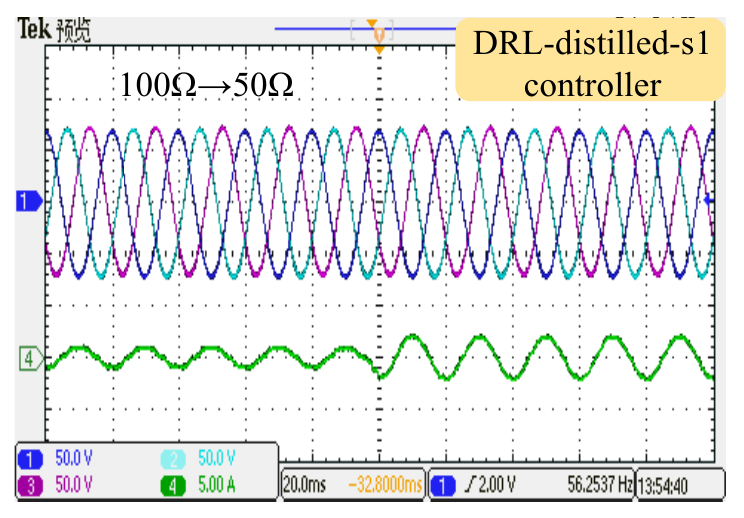}
       \caption{}
        \label{fig:exp:b}
    \end{subfigure}
    \hfill
    \begin{subfigure}[b]{0.245\linewidth}
        \centering
        \includegraphics[width=\textwidth]{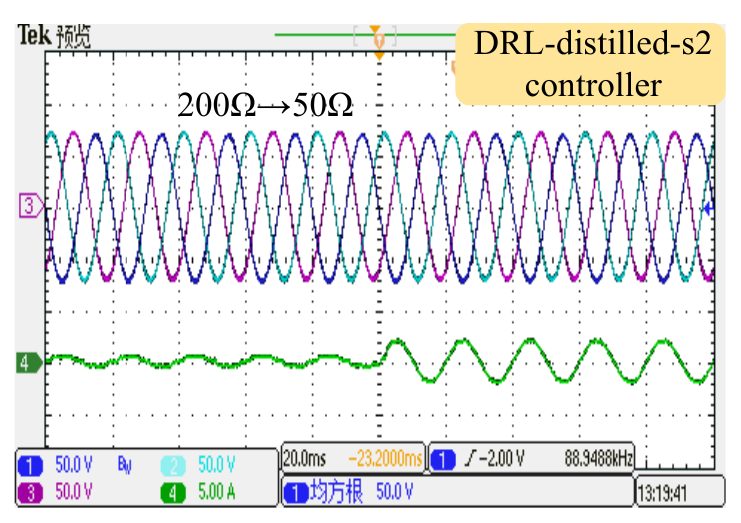}
        \caption{}
        \label{fig:exp:c}
    \end{subfigure}
    \hfill
    \begin{subfigure}[b]{0.245\linewidth}
        \centering
        \includegraphics[width=\textwidth]{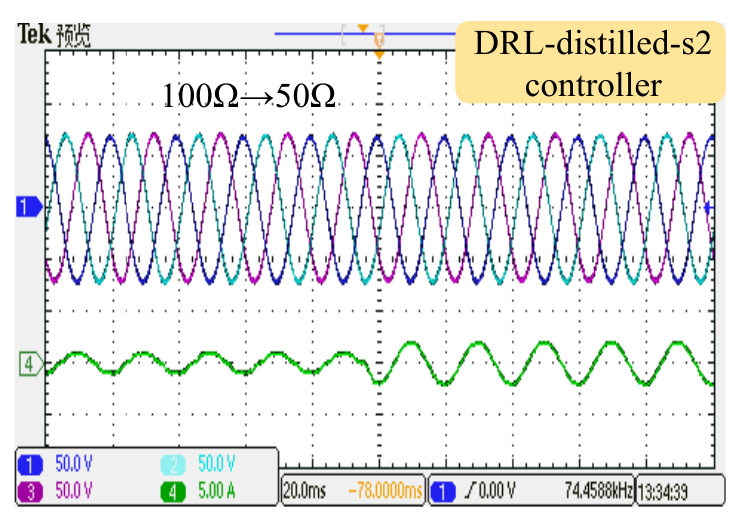}
        \caption{}
        \label{fig:exp:d}
    \end{subfigure}

    \vspace{0.8em} % 两行之间的垂直间距
    
    % --- 第二行 (e) 到 (h) ---
    \begin{subfigure}[b]{0.245\linewidth}
        \centering
        \includegraphics[width=\textwidth]{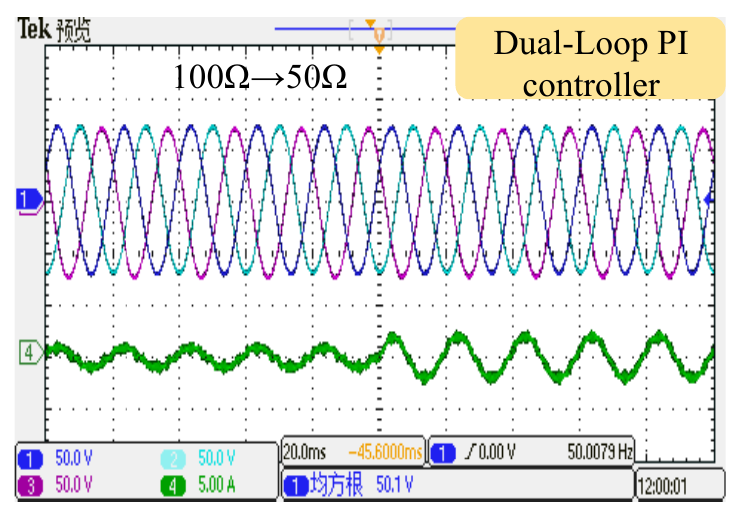}
       \caption{}
        \label{fig:exp:e}
    \end{subfigure}
    \hfill
    \begin{subfigure}[b]{0.245\linewidth}
        \centering
        \includegraphics[width=\textwidth]{paper3/images/PI-2.pdf}
        \caption{}
        \label{fig:exp:f}
    \end{subfigure}
    \hfill 
    \begin{subfigure}[b]{0.245\linewidth}
        \centering
        \includegraphics[width=\textwidth]{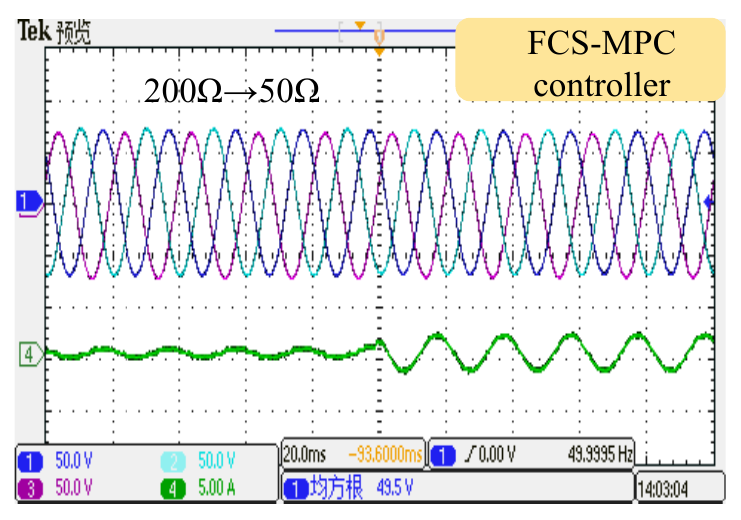}
        \caption{}
        \label{fig:exp:g}
    \end{subfigure}
    \hfill
    \begin{subfigure}[b]{0.245\linewidth}
        \centering
        \includegraphics[width=\textwidth]{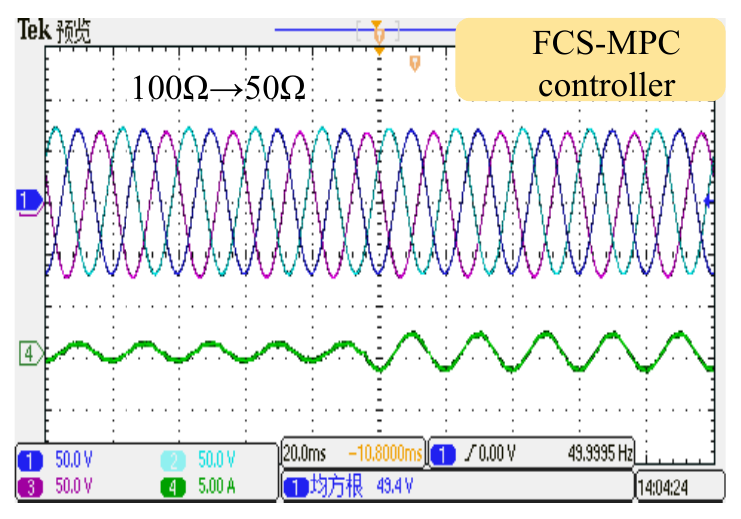}
        \caption{}
        \label{fig:exp:h}
    \end{subfigure}
    
     \caption{Hardware experimental comparison of controllers under different load steps.}
      \label{exam}
\end{figure*}
\begin{table}[htbp]
  \centering
  \caption{Quantitative Ablation Comparison of Policy Distillation}
  \label{tab:distill}
  
  % 使用 resizebox 强制表格适应单栏宽度
  \resizebox{\columnwidth}{!}{
    \begin{tabular}{@{}lccccc@{}}
      \toprule
      \textbf{Method} & 
      \textbf{Params.} & 
      \makecell{\textbf{Infer.} \\ \textbf{FLOPs}} & 
      \makecell{\textbf{Est. Time} \\ \textbf{($\mu$s)}} & % 删除了 \tnote{*}
      \makecell{\textbf{Comp.} \\ \textbf{Ratio}} & 
      \makecell{\textbf{Distill.} \\ \textbf{Loss}} \\
      \midrule
      
      Proposed DRL & 13442 & $\sim$26.4 k & 33.0 & 1 & / \\
      
      DRL-Distilled-S1 & 2687 & $\sim$5.2 k & 6.5 & 5 & 0.025/0.069 \\
      
      DRL-Distilled-S2 & 487 & $\sim$0.9 k & 1.1 & 26.7 & 0.024/0.079 \\
      \bottomrule
    \end{tabular}
  }
\end{table}

\begin{figure}[thpb]
  \centering
  \includegraphics[width=1.0\linewidth]{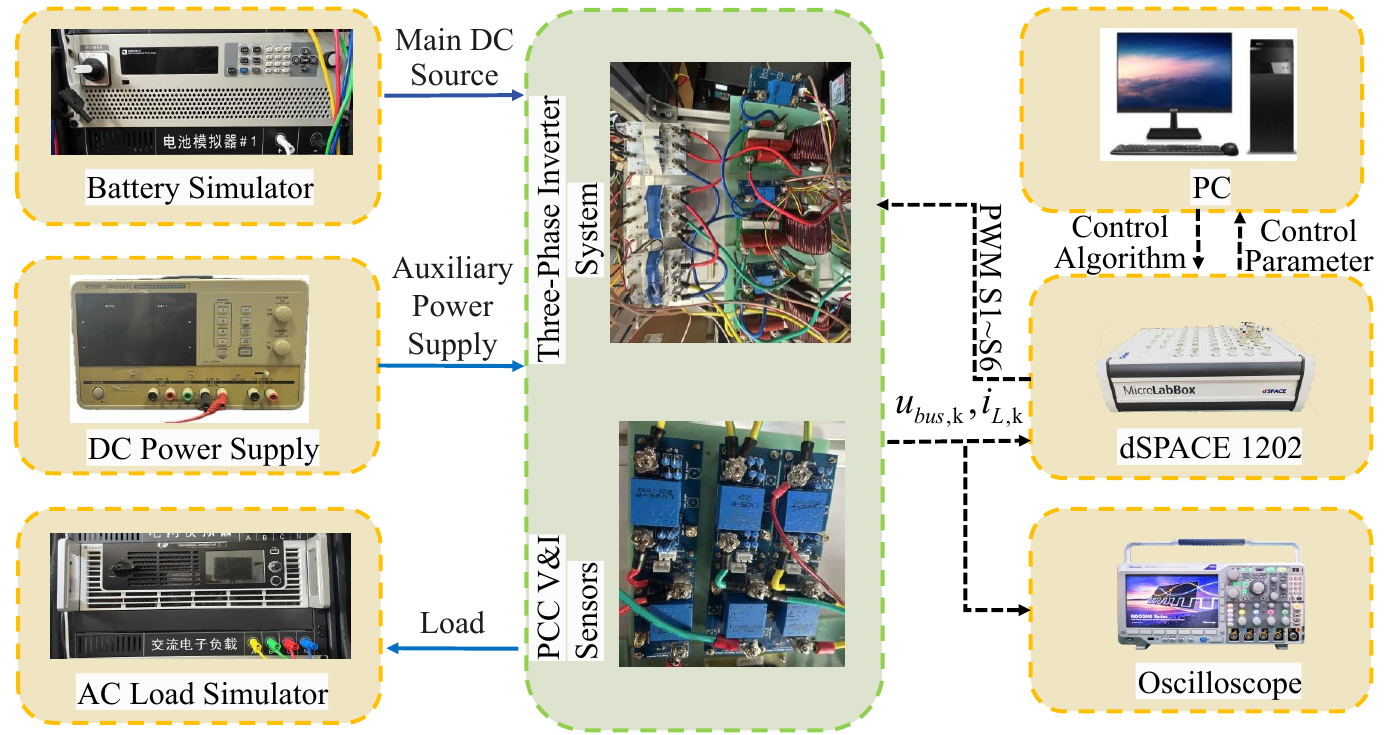} % Placeholder
  \caption{Experimental setup.}
    \label{fig:5}

\end{figure}
 \vspace{-10pt}

\subsection{Experimental Results}
To verify the practical feasibility of the proposed lightweight control strategy, a hardware experimental platform was built as shown in Fig. \ref{fig:5}.The power stage core of this platform is a three-phase voltage source inverter, which constitutes the main circuit together with a programmable DC power supply and an AC electronic load. The control system adopts an architecture comprising a dSPACE 1202 MicroLabBox and a host computer; the former is responsible for real-time low-level control and signal acquisition, while the latter runs the SAC algorithm within the MATLAB environment for policy inference.

On the hardware platform, the disturbance scenario identical to Case 1 in the simulation was reproduced, and key waveforms were captured using a digital oscilloscope. To ensure the reproducibility of experimental results and the fairness of the comparative analysis, all comparative tests were conducted on the exact same hardware platform and under identical external conditions, ensuring that the performance differences between control strategies originate solely from the algorithms themselves. Specific circuit design parameters are detailed in Table \ref{tab:circuit_params}.

Table \ref{tab:case1_comparison} presents the quantitative experimental comparison of performance metrics for various controllers under the Case 1 load step scenario. Fig. \ref{exam} shows the hardware experimental waveforms of the DRL-Distilled controllers under different load steps. From the waveforms, it is evident that the student networks exhibit rapid transient recovery capabilities and high steady-state waveform quality. Furthermore, to verify the real-time deployment capability of the proposed policy distillation strategy, Table \ref{tab:execution_time} provides a detailed comparison of the total execution time and the average time consumption per control cycle for different controllers. The policy-distilled DRL controllers outperform traditional PI and FCS-MPC in terms of total execution time, and the single-cycle time consumption is maintained at the microsecond level. This strongly confirms that the proposed algorithm imposes a low computational load, satisfying the real-time deployment requirements of systems.

\begin{table}[htbp]
    \caption{Comparative analysis of experimental results under case 1}
    \label{tab:case1_comparison}
    \centering
    \setlength{\tabcolsep}{4pt} % 2. 核心修改：将列间距减小为 3pt (默认约 6pt)
    \begin{tabular}{lccccc}
        \toprule
       {Controller} & \makecell{{Load} \\ ($\Omega$)} & \makecell{{Voltage} \\ THD($\%$)} & \makecell{{Current} \\ THD($\%$)} & \makecell{{SSE} \\ ($\%$)} & \makecell{{Unbalance} \\ ($\%$)} \\
        \midrule
        \multirow{2}{*}{\centering PI} 
        & $100 \rightarrow 50$ & $0.7628$ & $5.2778$ & $0.1810$ & $0.4350$ \\
        & $200 \rightarrow 50$ & $0.7740$ & $5.4105$ & $0.1921$ & $0.2881$ \\
        \cmidrule{1-6}
        \multirow{2}{*}{\centering FCS-MPC} 
        & $100 \rightarrow 50$ & $0.7779$ & $4.2242$ & $0.1968$ & $0.2952$ \\
        & $200 \rightarrow 50$ & $0.8251$ & $4.0401$ & $0.5421$ & $0.8132$ \\
        \cmidrule{1-6}
        \multirow{2}{*}{\centering DRL-S1} 
        & $100 \rightarrow 50$ & $0.5526$ & $3.2379$ & $0.4814$ & $0.7221$ \\
        & $200 \rightarrow 50$ & $0.5642$ & $3.2580$ & $0.3880$ & $0.5819$ \\
        \cmidrule{1-6}
        \multirow{2}{*}{\centering DRL-S2} 
        & $100 \rightarrow 50$ & $0.5649$ & $3.3340$ & $0.4241$ & $0.6362$ \\
        & $200 \rightarrow 50$ & $0.5738$ & $3.3807$ & $0.4371$ & $0.6556$ \\
        \bottomrule
    \end{tabular}
      \vspace{-0.5em} 
\end{table}

\begin{table}[htbp]
  \centering
  \small
  \renewcommand{\arraystretch}{0.95}
  \setlength{\tabcolsep}{5pt}
  \caption{Comparison of execution time for different controllers}
  \label{tab:execution_time}
  \begin{tabular}{@{}l c c@{}}
    \toprule
    \makecell[c]{\textbf{Controller}} & 
    \makecell[c]{\textbf{Total time} \textbf{(s)}} & 
    \makecell[c]{\textbf{Time of each} \\\textbf{control cycle ($\mu$s)}} \\
    \midrule
    Dual-loop PI  & 11.256 & 2.849 \\
    FCS-MPC  & 43.923 & 12.087 \\
    DRL-Distilled-S1  & 3.808 & 1.215 \\
    DRL-Distilled-S2  & 3.753 & 1.206 \\
    \bottomrule
  \end{tabular}
  \vspace{-0.5em} 
\end{table}
 \vspace{-10pt}

\section{Conclusion}

This paper proposes a model-free deep reinforcement learning control strategy for voltage source inverters. By introducing a hybrid reward mechanism based on error energy, the proposed method effectively enhances training convergence and control accuracy. Furthermore, policy distillation techniques are employed to compress the complex model into a lightweight network, thereby resolving computational bottlenecks associated with hardware deployment. Simulation and experimental results validate that the proposed strategy exhibits superior dynamic performance and robustness compared to traditional PI control and FCS-MPC under load steps and parameter mismatches, while its computational efficiency fully satisfies the real-time requirements of systems.
 \vspace{-0.5em} 
\bibliographystyle{IEEEtran}
\bibliography{ref}

\begin{thebibliography}{10}
\providecommand{\url}[1]{#1}
\csname url@samestyle\endcsname
\providecommand{\newblock}{\relax}
\providecommand{\bibinfo}[2]{#2}
\providecommand{\BIBentrySTDinterwordspacing}{\spaceskip=0pt\relax}
\providecommand{\BIBentryALTinterwordstretchfactor}{4}
\providecommand{\BIBentryALTinterwordspacing}{\spaceskip=\fontdimen2\font plus
\BIBentryALTinterwordstretchfactor\fontdimen3\font minus \fontdimen4\font\relax}
\providecommand{\BIBforeignlanguage}[2]{{%
\expandafter\ifx\csname l@#1\endcsname\relax
\typeout{** WARNING: IEEEtran.bst: No hyphenation pattern has been}%
\typeout{** loaded for the language `#1'. Using the pattern for}%
\typeout{** the default language instead.}%
\else
\language=\csname l@#1\endcsname
\fi
#2}}
\providecommand{\BIBdecl}{\relax}
\BIBdecl

\bibitem{Wang2019Harmonic}
X.~Wang and F.~Blaabjerg, ``Harmonic stability in power electronic-based power systems: {Concept}, modeling, and analysis,'' \emph{IEEE Transactions on Smart Grid}, vol.~10, no.~3, pp. 2858--2870, May 2019.

\bibitem{Buso2022Digital}
S.~Buso and P.~Mattavelli, \emph{Digital Control in Power Electronics}, ser. Synthesis Lectures on Power Electronics.\hskip 1em plus 0.5em minus 0.4em\relax Cham, Switzerland: Morgan \& Claypool Publishers, 2022.

\bibitem{Hou2013From}
Z.-S. Hou and Z.~Wang, ``From model-based control to data-driven control: Survey, classification and perspective,'' \emph{Information Sciences}, vol. 235, pp. 3--35, Jun 2013.

\bibitem{Chen2016Disturbance}
W.-H. Chen, J.~Yang, L.~Guo, and S.~Li, ``Disturbance-observer-based control and related methods—an overview,'' \emph{IEEE Transactions on Industrial Electronics}, vol.~63, no.~2, pp. 1083--1095, Feb 2016.

\bibitem{Niu2024New}
X.~Niu, C.~Zhang, Y.~Qu, X.~Wang, and H.~Li, ``A new composite control strategy for large-signal stabilization of constant power loads in islanded {AC} microgrids,'' \emph{IEEE Transactions on Smart Grid}, vol.~15, no.~5, pp. 4407--4423, Sep 2024.

\bibitem{Zhao2021Observer}
Y.~Zhao, X.~Li, and S.~Song, ``Observer-based sliding mode control for stabilization of mismatched disturbance systems with or without time delays,'' \emph{IEEE Transactions on Systems, Man, and Cybernetics: Systems}, vol.~51, no.~12, pp. 7337--7345, Dec 2021.

\bibitem{Lin2025Robust}
Y.-S. Lin, K.-Z. Liu, P.-F. Liu, and M.-S. Yang, ``A robust finite-time control strategy for a three-phase inverter with {LC} filter under uncertain disturbances,'' \emph{IEEE Transactions on Circuits and Systems I: Regular Papers}, pp. 1--12, Aug 2025, early Access.

\bibitem{Khalilzadeh2021Model-Free}
M.~Khalilzadeh, S.~Vaez-Zadeh, J.~Rodriguez, and R.~Heydari, ``Model-free predictive control of motor drives and power converters: A review,'' \emph{IEEE Access}, vol.~9, pp. 105\,733--105\,747, Jul 2021, open Access under Creative Commons License; Cited by 125; Full Text Views: 6199; Classic review on MFPC for motor drives and power converters.

\bibitem{Li2022Revisit}
W.~Li, H.~Yuan, S.~Li, and J.~Zhu, ``A revisit to model-free control,'' \emph{IEEE Transactions on Power Electronics}, vol.~37, no.~12, Dec 2022.

\bibitem{Wei2025ModelFree}
Y.~Wei, Y.~Chen, C.~Garcia, Z.~Yin, F.~Wang, and J.~Rodriguez, ``Model-free predictive control for {PMSM} drives using ultra-local with {Poincaré} mapping causality,'' \emph{IEEE Transactions on Industrial Electronics}, pp. 1--12, Oct 2025.

\bibitem{Govinda2025Survey}
S.~Govinda, B.~Brik, and S.~Harous, ``A survey on deep reinforcement learning applications in autonomous systems: Applications, open challenges, and future directions,'' \emph{IEEE Transactions on Intelligent Transportation Systems}, vol.~26, no.~7, pp. 11\,088--11\,113, Jul 2025.

\bibitem{She2023Fusion}
B.~She, F.~Li, H.~Cui, J.~Zhang, and R.~Bo, ``Fusion of microgrid control with model-free reinforcement learning: Review and vision,'' \emph{IEEE Transactions on Smart Grid}, vol.~14, no.~4, pp. 3232--3245, Jul 2023.

\bibitem{Mohammadi2022Review}
E.~Mohammadi, M.~Alizadeh, M.~Asgarimoghaddam, X.~Wang, and M.~G. Simões, ``A review on application of artificial intelligence techniques in microgrids,'' \emph{IEEE Journal of Emerging and Selected Topics in Industrial Electronics}, vol.~3, no.~4, pp. 878--890, Oct 2022.

\bibitem{Rajamallaiah2025DRL}
A.~Rajamallaiah, S.~V.~K. Naresh, Y.~Raghuvamsi, S.~Manmadharao, K.~Bingi, and A.~R, ``Deep reinforcement learning for power converter control: A comprehensive review of applications and challenges,'' \emph{IEEE Open Journal of Power Electronics}, vol.~6, pp. 1769--1802, Oct 2025.

\bibitem{Cui2025Domain}
C.~Cui, Z.~Fan, T.~Yang, P.~Lin, C.~Gong, and C.~Zhang, ``Domain adaptation-based transfer learning for {DRL} control implementation of {DC} microgrids,'' \emph{IEEE Transactions on Industrial Electronics}, pp. 1--12, Jun 2025.

\bibitem{Zhou2024DRL}
L.~Zhou, C.~Zhang, C.~Cui, P.~Lin, and X.~Dong, ``A {DRL}-based parameter self configuration mechanism of nonsmooth control for autonomous {DC} microgrids feeding constant power loads,'' \emph{IEEE Journal of Emerging and Selected Topics in Power Electronics}, vol.~12, no.~1, pp. 641--650, Feb 2024.

\bibitem{Wan2025Stability}
Y.~Wan and Q.~Xu, ``Stability-guided reinforcement learning control for power converters: A lyapunov approach,'' \emph{IEEE Transactions on Industrial Electronics}, vol.~72, no.~7, pp. 7553--7553, Jul 2025.

\bibitem{Li2020EdgeAI}
E.~Li, L.~Zeng, Z.~Zhou, and X.~Chen, ``Edge {AI}: On-demand accelerating deep neural network inference via edge computing,'' \emph{IEEE Transactions on Wireless Communications}, vol.~19, no.~1, pp. 447--457, Jan 2020.

\bibitem{Chen2024RLReview}
P.~Chen, J.~Zhao, K.~Liu, J.~Zhou, K.~Dong, and Y.~Li, ``A review on the applications of reinforcement learning control for power electronic converters,'' \emph{IEEE Transactions on Industry Applications}, vol.~60, no.~6, pp. 8430--8450, Nov 2024.

\bibitem{Romero2015FitNets}
A.~Romero, N.~Ballas, S.~E. Kahou, A.~Chassang, C.~Gatta, and Y.~Bengio, ``Fitnets: Hints for thin deep nets,'' in \emph{Proceedings of the 3rd International Conference on Learning Representations (ICLR)}, 2015.

\bibitem{Yang2022Focal}
Z.~Yang, Z.~Li, X.~Jiang, Y.~Gong, Z.~Yuan, D.~Zhao, and C.~Yuan, ``Focal and global knowledge distillation for detectors,'' in \emph{Proceedings of the IEEE/CVF Conference on Computer Vision and Pattern Recognition (CVPR)}, 2022, pp. 4643--4652.

\bibitem{xin2006engineering}
T.~Xin, Z.~Xiangjun, and T.~Chunming, ``The engineering design and optimization of inverter output {LCR} filter in parallel active power filter,'' in \emph{2006 International Conference on Power System Technology}.\hskip 1em plus 0.5em minus 0.4em\relax IEEE, 2006, pp. 1--6.

\bibitem{zeng2025multi}
Y.~Zeng \emph{et~al.}, ``Multi-agent soft actor-critic aided active disturbance rejection control of {DC} solid-state transformer,'' \emph{IEEE Transactions on Industrial Electronics}, vol.~72, no.~1, pp. 492--503, Jan 2025.

\bibitem{pei2023multitask}
Y.~Pei, J.~Zhao, Y.~Yao, and F.~Ding, ``Multi-task reinforcement learning for distribution system voltage control with topology changes,'' \emph{IEEE Transactions on Smart Grid}, vol.~14, no.~3, pp. 2481--2484, May 2023.

\bibitem{oshnoei2024grid}
A.~Oshnoei, H.~Sorouri, S.~Oshnoei, R.~Teodorescu, and F.~Blaabjerg, ``Grid impedance shaping for {Grid-Forming} inverters: A {Soft} {Actor-Critic} deep reinforcement learning algorithm,'' in \emph{2024 IEEE 10th International Power Electronics and Motion Control Conference (IPEMC2024-ECCE Asia)}.\hskip 1em plus 0.5em minus 0.4em\relax Chengdu, China: IEEE, 2024, pp. 4935--4939.

\bibitem{wang2025scaling}
K.~Wang, M.~Bortkiewicz, I.~Javali, T.~Trzci{\'n}ski, and B.~Eysenbach, ``1000 layer networks for self-supervised {RL}: Scaling depth can enable new goal-reaching capabilities,'' in \emph{Advances in Neural Information Processing Systems (NeurIPS)}, 2025.

\bibitem{Cui2025DomainAdaptation}
C.~Cui, Z.~Fan, T.~Yang, P.~Lin, C.~Gong, and C.~Zhang, ``Domain adaptation-based transfer learning for {DRL} control implementation of {DC} microgrids,'' \emph{{IEEE} Transactions on Industrial Electronics}, vol.~72, no.~12, pp. 14\,344--14\,355, Dec. 2025.

\bibitem{lou2021physics}
Q.~Lou, X.~Meng, and G.~E. Karniadakis, ``Physics-informed neural networks for solving forward and inverse flow problems via the boltzmann-bgk formulation,'' \emph{Journal of Computational Physics}, vol. 447, p. 110676, 2021.

\bibitem{monshizadeh2022nonlinear}
N.~Monshizadeh, F.~Mancilla-David, R.~Ortega, and R.~Cisneros, ``Nonlinear stability analysis of the classical nested {PI} control of voltage sourced inverters,'' \emph{IEEE Control Systems Letters}, vol.~6, pp. 1442--1447, 2022.

\bibitem{cortes2009model}
P.~Cort{\'e}s, G.~Ortiz, J.~I. Yuz, J.~Rodr{\'\i}guez, S.~Vazquez, and L.~G. Franquelo, ``Model predictive control of an inverter with output {LC} filter for {UPS} applications,'' \emph{IEEE Transactions on Industrial Electronics}, vol.~56, no.~6, pp. 1875--1883, Jun 2009.

\end{thebibliography}
\vspace{-2cm}
\begin{IEEEbiography}[{\includegraphics[width=1in,height=1.25in,clip,keepaspectratio]{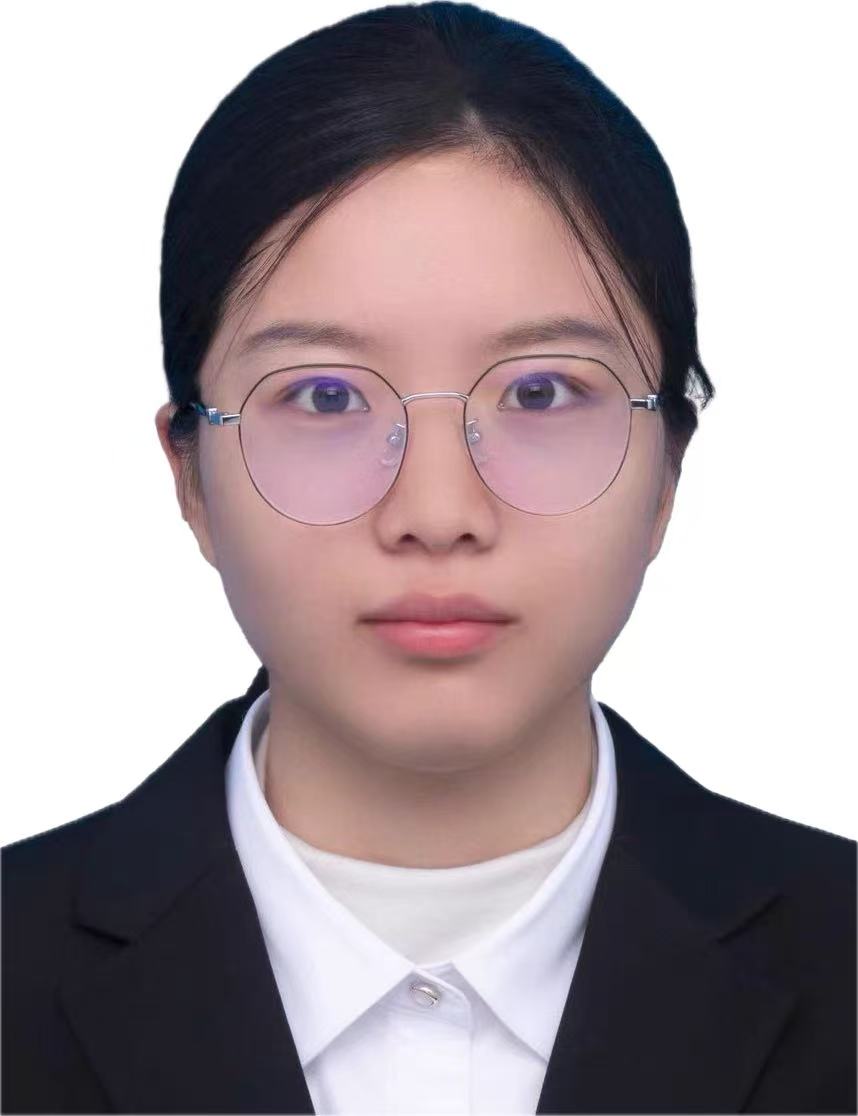}}]
{Yang Yang} received the B.E. degree in Electrical Engineering from Nanjing University of Information Science \& Technology, Nanjing, China, in 2022. He is currently working toward the M.S. degree in Automation Engineering at Shanghai University of Electric Power, Shanghai, China.He is a member of Intelligent Autonomous Systems Laboratory. His research focuses on the application of artificial intelligence in power electronics systems.
\end{IEEEbiography}

\begin{IEEEbiography}[{\includegraphics[width=1in,height=1.25in,clip,keepaspectratio]{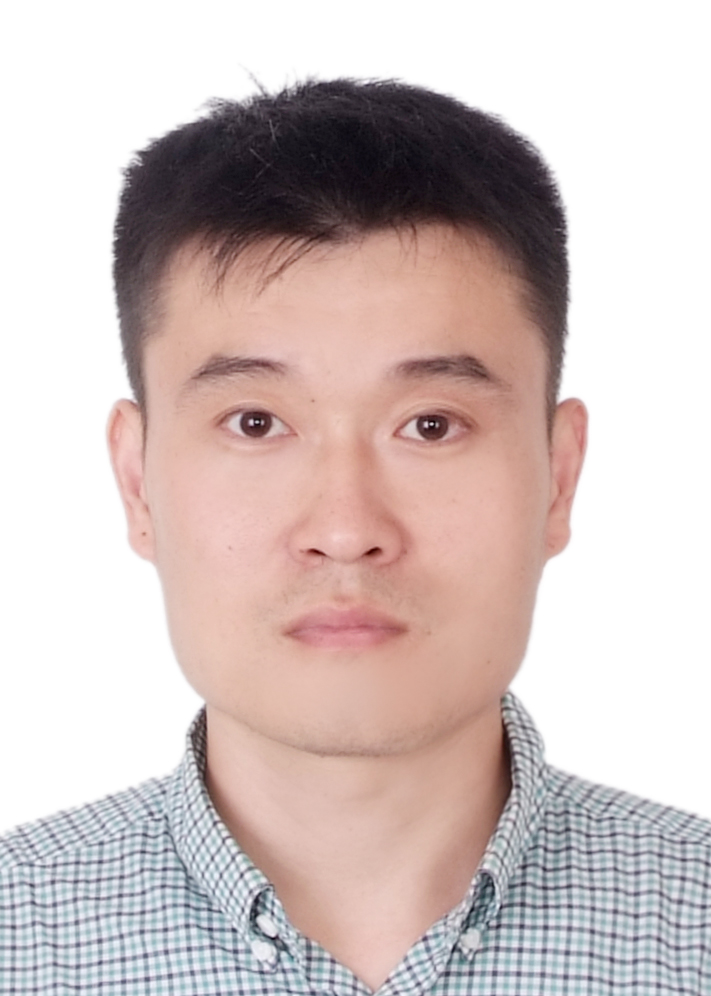}}]
{Chengang Cui} (Member, IEEE) received the B.E. degree in automation engineering from Jilin University, China, in 2004, the Ph.D. degree in control theory and control from Zhejiang University, China, in 2010. He worked at Shanghai Institute for Advanced Studies, Chinese Academy of Sciences, engaged in energy management and optimal scheduling from 2012 to 2015. He has been with the School of Automation, Shanghai University of Electric Power, where he is currently a associate professor. His research interests cover the control.
\end{IEEEbiography}

\begin{IEEEbiography}[{\includegraphics[width=1in,height=1.25in,clip,keepaspectratio]{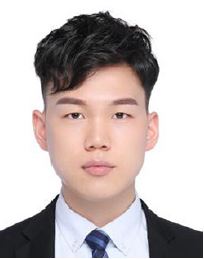}}]
{Xitong Niu} (Student Member, IEEE) received the
M.Sc. degree in electrical engineering from the
Shanghai University of Electric Power, Shanghai,
China, in 2018, where he is currently pursuing
the Ph.D. degree in electrical engineering with
the School of Electrical Engineering. He was with
Unitech Embedded and KUKA Company from 2018
to 2023, focusing on power electronic systems. His
research interests include advanced algorithm in
power electronic systems and microgrids.
\end{IEEEbiography}
\vspace{-1cm}

\begin{IEEEbiography}[{\includegraphics[width=1in,height=1.25in,clip,keepaspectratio]{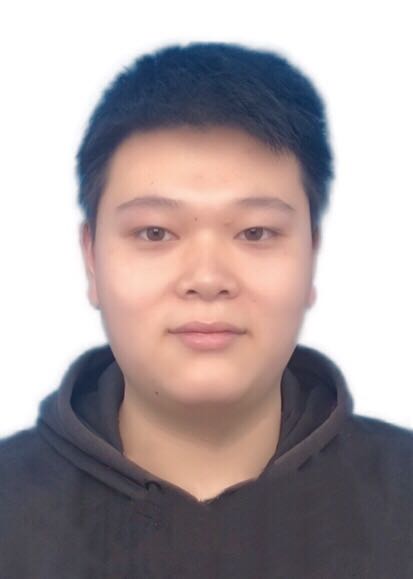}}]
{Jiaming Liu} received the B.E. degree in Electrical Engineering from Nanjing University of Information Science \& Technology, Nanjing, China, in 2022. He is currently working toward the M.S. degree in Automation Engineering at Shanghai University of Electric Power, Shanghai, China.He is a member of Intelligent Autonomous Systems Laboratory. His research focuses on the application of Large Language Model.
\end{IEEEbiography}
\begin{IEEEbiography}[{\includegraphics[width=1in,height=1.25in,clip,keepaspectratio]{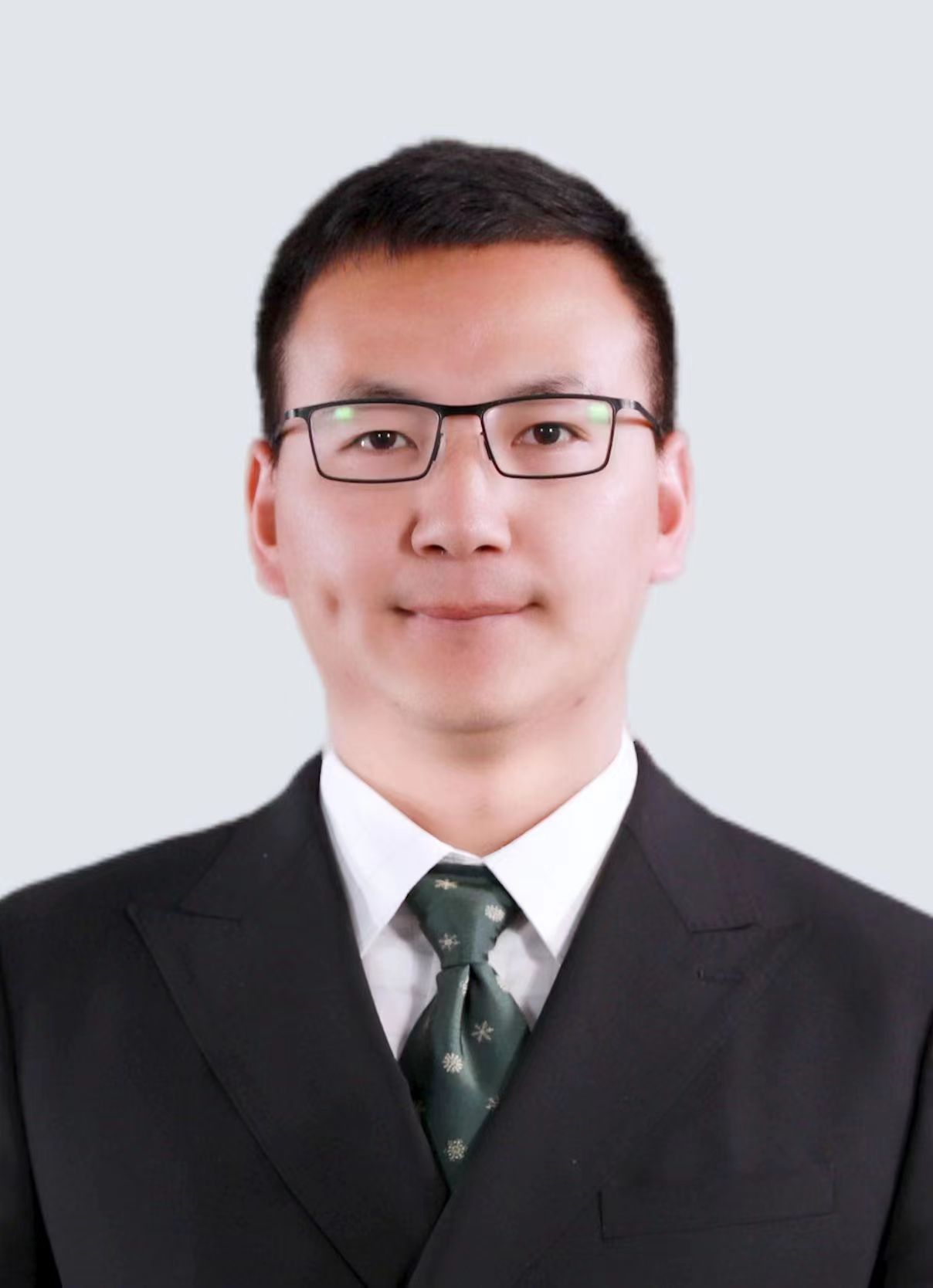}}]
{Chuanlin Zhang} (Senior Member, IEEE) received the B.S. degree in mathematics and the Ph.D. degree in control theory and control engineering from the School of Automation, Southeast University, Nanjing, China, in 2008 and 2014, respectively. He was a Visiting Ph.D. Student with the Department of Electrical and Computer Engineering, University of Texas at San Antonio, USA, from 2011 to 2012; a Visiting Scholar with the Energy Research Institute, Nanyang Technological University, Singapore, from 2016 to 2017. Since 2014, he has been with the College of Automation Engineering, Shanghai University of Electric Power, Shanghai, where he is currently a Professor. His research interests include nonlinear system control theory and applications for power systems.
\end{IEEEbiography}
\end{document}